\documentclass[preprint]{aastex63}

\usepackage{color}
\usepackage{multirow}
\usepackage{threeparttable}
\usepackage{ulem}

\begin{document}
\shortauthors{Jiao, Wang et al.}

\title{Fragmentation of the High-mass ``Starless'' Core G10.21-0.31: a Coherent Evolutionary Picture for Star Formation}

\author{Wenyu Jiao}
\affiliation{Kavli Institute for Astronomy and Astrophysics, Peking University, Haidian District, Beijing 100871, People’s Republic of China}
\affiliation{Department of Astronomy, School of Physics, Peking University, Beijing, 100871, People's Republic of China}
\author{Ke Wang}
\affiliation{Kavli Institute for Astronomy and Astrophysics, Peking University, Haidian District, Beijing 100871, People’s Republic of China}
\author{Thushara G.S. Pillai}
\affiliation{Institute for Astrophysical Research, Boston University, 725 Commonwealth Avenue, Boston MA, 02215, USA}
\author{Tapas Baug}
\affiliation{S. N. Bose National Centre for Basic Sciences, Block-JD, Sector-III, Salt Lake City, Kolkata 700106, India}
\author{Siju Zhang}
\affiliation{Kavli Institute for Astronomy and Astrophysics, Peking University, Haidian District, Beijing 100871, People’s Republic of China}
\author{Fengwei Xu}
\affiliation{Kavli Institute for Astronomy and Astrophysics, Peking University, Haidian District, Beijing 100871, People’s Republic of China}
\affiliation{Department of Astronomy, School of Physics, Peking University, Beijing, 100871, People's Republic of China}

\correspondingauthor{Ke Wang}
\email{kwang.astro@pku.edu.cn}












\begin{abstract}

G10.21-0.31 is a 70\,$\mu$m-dark high-mass starless core ($M>300\, \mathrm{M_{\odot}}$ within $r<0.15$ pc) identified in {\it Spitzer, Herschel}, and APEX continuum surveys, and is believed to harbor the initial stages of high-mass star formation. We present ALMA and SMA observations to resolve the internal structure of this promising high-mass starless core. Sensitive high-resolution ALMA 1.3 mm dust continuum emission reveals three cores of mass ranging 11-18 $\mathrm{M_{\odot}}$, characterized by a turbulent fragmentation. Core 1, 2, and 3 represent a coherent evolution at three different evolutionary stages, characterized by outflows (CO, SiO), gas temperature ($\mathrm{H_2CO}$), and deuteration ($\mathrm{N_2D^+/N_2H^+}$). We confirm the potential to form high-mass stars in G10.21 and explore the evolution path of high-mass star formation. Yet, no high-mass prestellar core is present in G10.21. 
This suggests a dynamical star formation where cores grow in mass over time.

\end{abstract}



\section{Introduction}

\par High-mass stars play essential roles in the evolution of their host galaxy via their radiation, stellar wind and supernovae events. However, compared with their low-mass counterparts, the formation mechanism of high-mass stars is poorly understood \citep{Tan_2014,Motte_2018}.
\par There are two mainstream models describing the forming process of high-mass stars: the monolithic collapse model \citep{McKee_2003} and the competitive accretion model \citep{Bonnell_2001}. In the monolithic collapse model, the final stellar mass is pre-assembled in a single high-mass turbulent core. Turbulence can effectively resist self-gravity until the accretion mass is large enough. So this model requires the existence of high-mass starless cores. In the competitive accretion model, high-mass stars begin as low-mass cores ($\sim1  \  \mathrm{M_{\odot}}$) and most of the stellar mass comes from the subsequent accretion process. Different from Bondi-Hoyle-Lyttleton-like core accretion model, recent studies also propose some new accretion mechanisms such as gravitationally driven cloud inflow (\citealt{Smith_2009,Hartmann_2012}) or supersonic turbulence driven inflow \citep{Padoan_2020}. However, both the mechanisms are yet to get enough observational evidence.
\par In order to distinguish the different forming processes of high-mass stars, it is important to probe the initial conditions towards the birthplace of high-mass star forming regions. Infrared-dark clouds (IRDCs), often considered as the cradle of massive stars, provide ideal laboratories to study the formation of high-mass stars \citep{Bergin_2007}. In recent years, many high-resolution and high-sensitivity observations have revealed the physical properties of the cores embedded in IRDCs, suggesting that most of the clumps in IRDCs have left the starless stage and begun star formation activities (e.g., \citealt{Zhang_2009,Zhang_2015,Wang_2011,Wang_2014,Sanhueza_2013,Sanhueza_2019,Svoboda_2019,Pillai_2019,Barnes_2021}). 
\par Blind surveys of continuum emission at far-infrared and  (sub)millimeter wavelengths towards the Galactic plane reveal the dense structures at different evolutionary stages (e.g., \citealt{Schuller_2009,Molinari_2010,Aguirre_2011,Eden_2017}), offering good data sets for us to search for the cradle of high-mass star forming regions. \citet{Yuan_2017} have selected 463 high-mass starless clump candidates based on The APEX Telescope Large Area Survey of the Galaxy (ATLASGAL) catalog \citep{Schuller_2009}. Compared with other high-mass starless clump catalogs (e.g., \citealt{Tackenberg_2012,Traficante_2015,Svoboda_2016}), this catalog has the largest sky coverage. Besides the commonly adopted infrared-dark criteria, this catalog also ruled out sources associated with reported star-forming indicators and performed extra visual checks on each source, which made it the most reliable high-mass starless clump candidate catalog. Detailed studies towards this sample can help us reveal the initial conditions of high-mass star forming regions. 


\subsubsection*{ G10.21-0.31}
\par
\begin{figure}[bt!]
    \centering
    \includegraphics[width=1.0\linewidth]{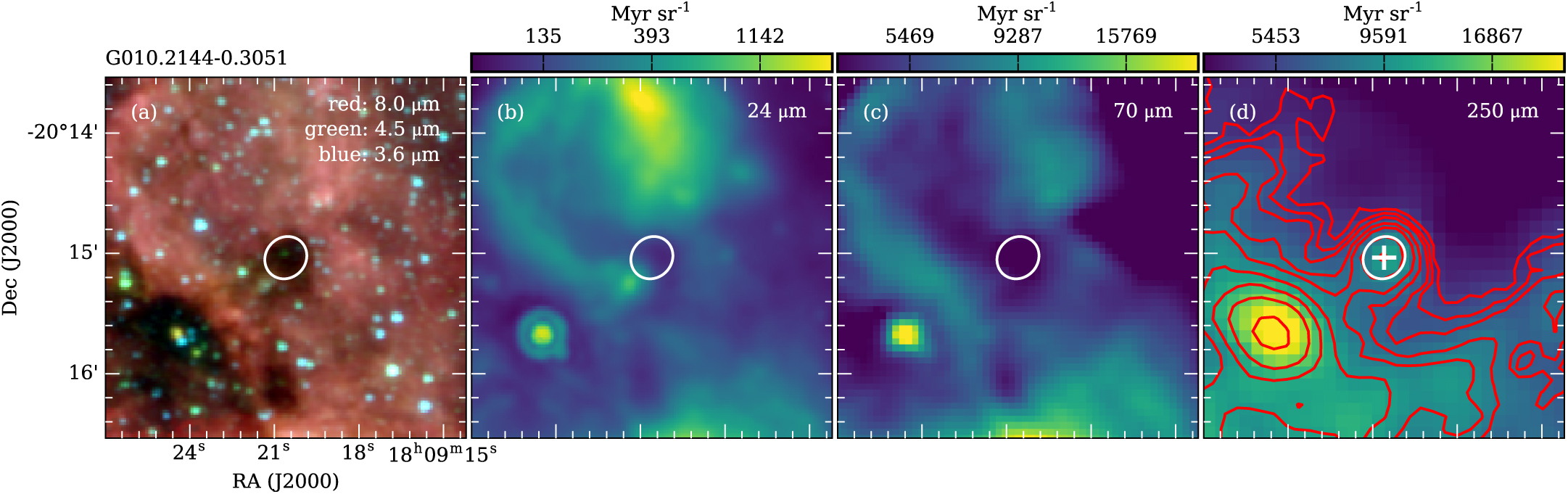}
    \caption{Overview of the high-mass starless core G10.21. Four panels show images at different wavelengths adapted from \citet{Yuan_2017}. (a) Three color image with emission at 8.0, 4.5, and 3.6 $\mu$m rendered in red, green, and blue, respectively. (b) Spitzer 24 $\mu$m emission. (c) Herschel 70 $\mu$m emission. (d) APEX 870 $\mu$m emission (red contours) with levels of [0.3, 0.4, 0.5, 0.7, 0.9, 1.3, 1.8, 2.5, 4, 7] Jy/beam overlaid on Herschel 250 $\mu$m map (color). The white cross marks the peak position of 870 $\mu$m source, and the white ellipse describes the source size based on the major and minor half-intensity axes. }
    \label{fig:overview of G10.21}
\end{figure}

 \citet{Yuan_2017} found twenty high-mass starless core candidates with an equivalent radius smaller than 0.15 pc. G10.21-0.31 (hereafter G10.21) is the second most massive source in the sample. Figure \ref{fig:overview of G10.21} shows the infrared emission of G10.21 at different wavelengths. This source is dark at near-infrared, mid-infrared, far-infrared up to 70 $\mu$m, and transitions to be bright at longer wavelengths (250/870 $\mu$m bright). Located at a distance of 3.1 kpc, this high-mass prestellar core candidate contains 314 $\mathrm{M_\odot}$ gas within an equivalent radius of 0.13 pc \citep{Yuan_2017}. The luminosity of the target is 667 $\mathrm{L_\odot}$, leading to a relatively low luminosity-to-mass ratio ($2.13 \ \mathrm{L_\odot/M_\odot} $, \citet{Yuan_2017}). The deuterium fraction of $\mathrm{NH_3}$ in this source is higher than $30 \%$, suggesting it is at a very young age \citep{Pillai_2007}. The dust temperature of G10.21 is 16.6 K under $36.4^{\prime\prime}$ resolution \citep{Yuan_2017} and the kinetic temperature derived from $\mathrm{NH_3}$ is 18.5 K under  $40^{\prime\prime}$ resolution \citep{Wienen_2012}. Because of its small size, low temperature, low luminosity-to-mass ratio, high mass and high deuterium fraction, this source is an ideal object to study the initial conditions of high-mass star formation. In addition, it is located at the edge of the HII region G010.232-0.301 \citep{Anderson_2011}, which may suffer strong environmental feedback. \citet{Zhang_2021} compared four pairs of high-mass starless clumps (near or far from HII regions) and found that this source is more likely to be affected by the surrounding HII regions. Hence, exploring the initial star formation processes of G10.21 is of great interest.
\par The paper is organized as follows. We describe the observations in Section \ref{sec:observaton} and show the main results in Section \ref{sec:result}. In Section \ref{sec:discussion}, we discuss potential bias from temperature correction and the gas fragmentation mechanism in G10.21. We also discuss the chemical evolution in star formation to investigate the possibility of chemical clocks. In addition, we study the potential to form high-mass stars and describe a simple possible evolutionary picture of G10.21. Finally, we give the summary of this work in Section \ref{sec:conclusion}.

\section{Observations and Data Reduction}
\label{sec:observaton}
\subsection{ALMA Band 6 Observations}
G10.21 was observed with ALMA in Band 6, as part of the Cold Cores with ALMA (CoCoA) survey (Project ID: 2016.1.01346.S, PI: Thushara G.S. Pillai). In this work, we combine the data of the 12 m main array and the 7m Atacama Compact Array (ACA). The observations had four wide spectral windows with a bandwidth of 1.875 GHz for 12 m array and 2 GHz for 7 m array centered on 216.89, 218.76, 231.12, 232.89 GHz. The spectral windows have a uniform channel width of 244.14 kHz ($\sim$ 0.32 km $\mathrm{s^{-1}}$ at 230 GHz). The phase reference center was R.A.\,(J2000) = 18:09:20.7\text { and Decl.\,}({J}2000) = -20:15:04.0. Quasars J1832-2039 and J1924-2914 were used for phase and bandpass calibration. For the 12 m array observations, the source was observed on March 31, 2017 and May 14, 2018. The maximum recoverable scale was $13.7^{\prime\prime}$ and the primary beam size was $25.9^{\prime\prime}$. The integration time was approximately 1.5 minutes. Titan was used for flux calibration. For the 7 m array observations, the source was observed in July-September, 2017. The maximum recoverable scale was $36.4^{\prime\prime}$ and the primary beam size was $44.4^{\prime\prime}$. The integration time was approximately 9 minutes. J1733-1304 was used for flux calibration.

\begin{table}
	\centering
	\caption{\centering{Summary of Spectral Line Information}}
	\label{line information}
	\begin{tabular}{m{1.5cm}m{4cm}m{3cm}m{1.5cm}m{4cm}m{2cm}cccccc} 
		\hline
		\hline
Line &  \ \ \ \ \ Transition &  Rest Freq. & $E_u/k$ & Velocity Resolution & Beam Size \\
 &  & $\ \ $ (GHz) & $ \ $(K)&$\ \ \ \ \ \ \  (\mathrm{km \ s^{-1}}$) & \ \ \  ($^{\prime\prime}\times ^{\prime\prime}$)  \\
\hline
$\mathrm{DCO^{+}}$ & \ \ \ \ \ $J$=$3-2$	&  216.112580 &	 20.74 &  \ \ \ \ \ \ \ \ \ \ 0.34 &    1.98$\times$1.14  \\
$\mathrm{SiO}$ & \ \ \ \ \ $J$=$5-4$	&  217.104980 &	 31.26 & \ \ \ \ \ \ \ \ \ \ 0.34 &   1.98$\times$1.14  \\
$\mathrm{DCN}$ & \ \ \ \ \ $J$=$3-2$	&  217.238530 &	 20.85  &\ \ \ \ \ \ \ \ \ \ 0.34 &   1.98$\times$1.14  \\
$\mathrm{H_2CO}$ & $J_{\mathrm{K}_{\mathrm{a}}, \mathrm{K}_{\mathrm{c}}}$=$3_{0,3}-2_{0,2}$	&  218.222192 &	20.96 &  \ \ \ \ \ \ \ \ \ \ 0.34 &   1.97$\times$1.12  \\
$\mathrm{H_2CO}$ & $J_{\mathrm{K}_{\mathrm{a}}, \mathrm{K}_{\mathrm{c}}}$=$3_{2,2}-2_{2,1}$	&  218.475632 &	68.09 &  \ \ \ \ \ \ \ \ \ \ 0.34 &    1.97$\times$1.12  \\
$\mathrm{H_2CO}$ & $J_{\mathrm{K}_{\mathrm{a}}, \mathrm{K}_{\mathrm{c}}}$=$3_{2,1}-2_{2,0}$	&  218.760066 &	68.11 &  \ \ \ \ \ \ \ \ \ \ 0.33 &    1.96$\times$1.12  \\
$\mathrm{CO}$ & \ \ \ \ \ $J$=$2-1$	&  230.538000 &	 16.60  &\ \ \ \ \ \ \ \ \ \ 0.32 &  1.83$\times$1.07  \\
$\mathrm{N_2D^{+}}$ & \ \ \ \ \ $J$=$3-2$	&  231.321828 &	 22.20 &  \ \ \ \ \ \ \ \ \ \ 0.32 &  1.82$\times$1.07  \\
\hline
	\end{tabular}
\end{table}

\par Data reduction was performed using CASA software package version 5.6.1 \citep{McMullin_2007}. For continuum data, we use the \emph{split} task to obtain line-free channels in each spectral window. The visibility data of 12 m array and 7 m array from the four spectral windows were combined in CASA using \emph{concat} task. The combined 12 + 7 m continuum visibility data was cleaned using \emph{tclean} task, with a Briggs's robust weighting of 0.5 and a cell size of $0.3^{\prime \prime}$.  The synthesized beam size is $1.7^{\prime \prime} \times 1.0^{\prime \prime} $ (0.03 pc $\times$ 0.02 pc at 3.1 kpc distance), with a position angle of $-76^{\circ}$. The RMS noise of the continuum image measured in an emission-free region is about 0.5 mJy/beam ($\sim$0.15 $\mathrm{M_{\odot}}$ with a 16.6 K dust temperature at 3.1 kpc). For spectral line, the combined 12 + 7 m line cube was cleaned using \emph{tclean} task after removing the continuum emission using \emph{uvcontsub} task, with a Briggs robust weighting of 0.5 and a cell size of $0.3^{\prime \prime}$. The threshold is 2$\sigma$=0.025 Jy and the maximum number of iteration (niter) is 10000. Multi-scale Clean is used for CO, SiO and $\mathrm{H_2CO}$ lines to better recover extended emission and scales are set to be [0, 5, 15]. We use auto-multithresh algorithm and the parameters are equal to the standard value for 12m + 7m combined data in official guides\footnote{see details in \url{ https://casaguides.nrao.edu/index.php?title=Automasking_Guide}}. The synthesized beam size of different lines is similar to the beam size of continuum image but has a small difference because of the frequency offset. We summarize the spectral line information used in our analysis in Table \ref{line information}. The typical noise level is $\sim$150 $\mathrm{mK}$ after converting the cube to brightness temperature units.

\subsection{SMA Observations}
G10.21 was observed by SMA (project ID: 2008A-S075, PI: Thushara G.S. Pillai). The observations were performed with the LO tuned at 225.4 GHz for $\mathrm{N_2D^+} \ (J=3-2)$ on August 24th, 2008 and 275.1 GHz for $\mathrm{N_2H^+}\ (J=3-2)$ on August 30th, 2008. Gain calibration was performed by periodic observations of quasars NRAO530 and $\text {J1911-201}$. Uranus was used for flux calibration. 3C454.3 was used for bandpass calibration. For $\mathrm{N_2D^+}$ observation, the on-source integration time was 98 minutes. The synthesized beam size was $6.4^{\prime \prime} \times 3.3^{\prime \prime} $ (0.10 pc $\times$ 0.05 pc), with a position angle of $23^{\circ}$. For $\mathrm{N_2H^+}$ observation, the on-source integration time was 24 minutes. The synthesized beam size was $9.8^{\prime \prime} \times 3.2^{\prime \prime} $ (0.15 pc $\times$ 0.05 pc), with a position angle of $-2^{\circ}$. To compare the two SMA cubes directly, we convolved the image to the same resolution and reprojected them into the same grid. The final synthesized beam was $9.8^{\prime \prime} \times 4.3^{\prime \prime} $ (0.15 pc $\times$ 0.06 pc), with a position angle of $-2^{\circ}$. The typical noise levels were 0.10 K and 0.03 K for $\mathrm{N_2H^+}$ and $\mathrm{N_2D^+}$ spectra, respectively.


\section{Results}
\label{sec:result}

\subsection{1.3 mm Compact Continuum Sources}
\label{sec:continuum}
\par
\begin{figure}[bt!]
    \centering
    \includegraphics[width=1.0\linewidth]{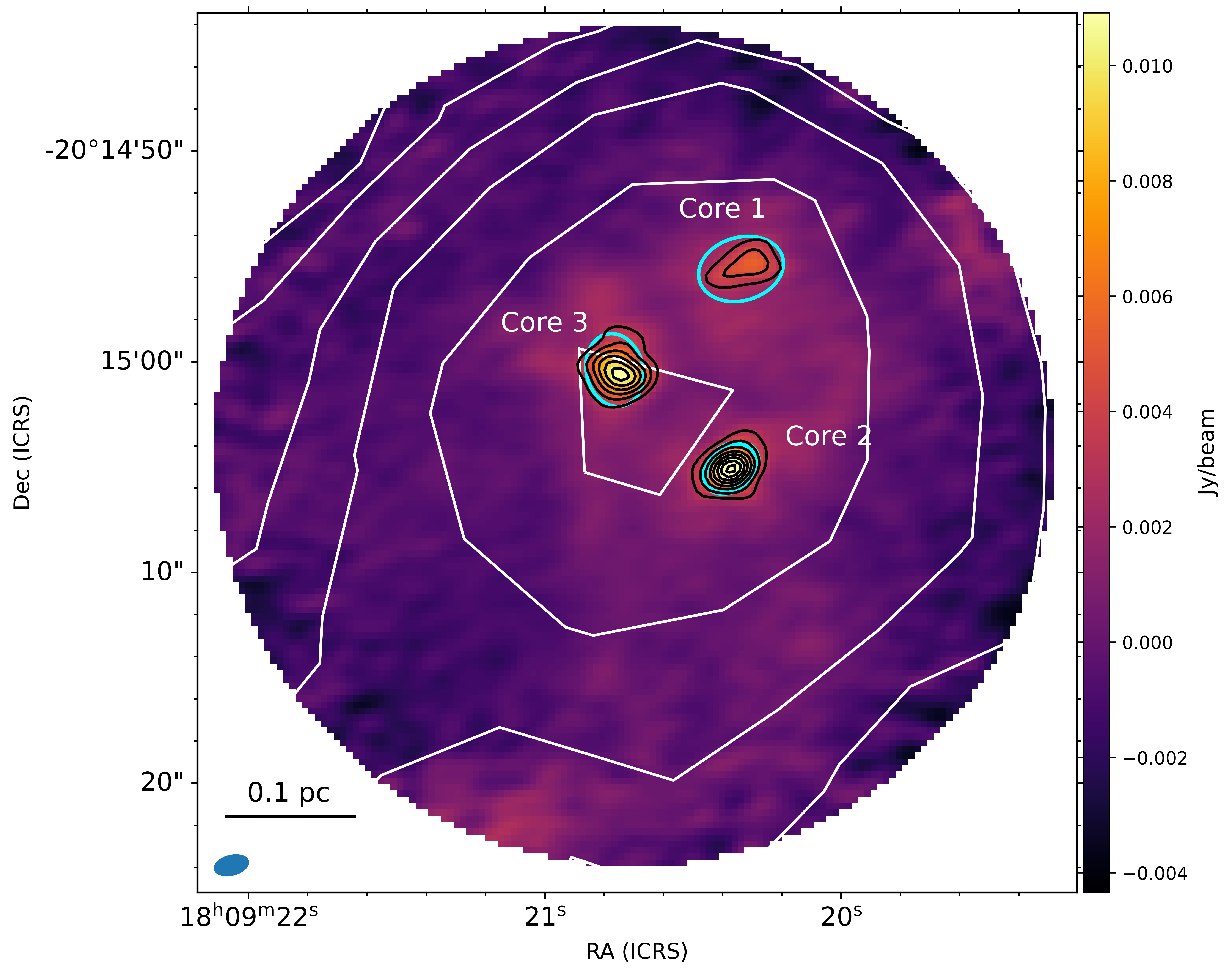}
    \caption{ALMA 1.3 mm continuum image of G10.21 shown in color after primary beam correction. The cyan ellipse represents the deconvolved image size of the cores. The white contours show emission at 870 $\mu$m from the ATLASGAL survey with the same levels in Figure \ref{fig:overview of G10.21}. The black contours show continuum emission at levels of [6, 9, 12, ...]$\sigma$, where $\sigma$ equals to $5 \times 10^{-4} \, \mathrm{Jy/beam}$. The blue ellipse in the bottom left marks synthesized beam.}
    \label{fig:continuum}
\end{figure}

  The 1.3 mm continuum map (centered at 224.92 GHz) is shown in Figure \ref{fig:continuum}. We use \emph{Dendrogram} algorithm\footnote{https://dendrograms.readthedocs.io/en/stable/} on the continuum image without primary beam correction to identify dense cores. The \emph{min\_value} is set to be 3$\sigma$, where $\sigma$ is the RMS noise of the continuum image. The \emph{min\_delta} is set to be 1.5$\sigma$ and the \emph{min\_npix} equals to 21 (The number of pixels within the beam area). To accurately derive the position, flux density, size information of these cores, we make use of \emph{imfit} function in CASA on the 1.3 mm continuum image after primary beam correction. We identified three compact cores in this source and the detailed observed properties are listed in Table \ref{Physical parameter}. All three cores are detected with high signal-to-noise ratios (S/N $\textgreater$6).

\par We estimate the core mass of three identified cores based on the dust continuum flux following the equation 

\begin{equation}
\centering
    M_{\mathrm{gas}}=\eta \frac{F_{\nu} d^{2}}{B_{\nu}(T_d) \kappa_{\nu}}
\end{equation}

where $M_{\mathrm{gas}}$ is the gas mass, $\eta=100$ is the gas-to-dust ratio, $F_{\nu}$ is the continuum flux at frequency $\nu$, $d$ is the kinetic distance of the source, $B_{\nu} \, (T_{d})$ is the Planck function at the dust temperature, and $\kappa_{\nu}=10 \, (\nu / 1.2 \ \mathrm{THz})^{\beta} \, \mathrm{cm}^{2} \, \mathrm{~g}^{-1}$ represents the dust opacity \citep{Hildebrand_1983}. In our calculation, we use the value of $T_{d}=16.6$ K derived for the whole region with $36.4^{\prime\prime}$ resolution in \citet{Yuan_2017} and adopt the dust opacity index $\beta=1.5$. If we adopt the value of $0.9 \mathrm{~cm}^{2} \mathrm{~g}^{-1}$ for dust opacity at 1.3 mm with a volume density of $10^6 \ \mathrm{cm^{-3}}$ density from \cite{OH94}, the derived gas masses from Core 1 to Core 3 will be 10.4, 12.6, 15.5 $\mathrm{M_{\odot}}$, leading to about 10\% difference to the final results.
\par The number density of each core is calculated assuming the sources to be spherically symmetric using the following equation:

\begin{equation}
    n_{\mathrm{H}_{2}}= \frac{M_{gas}}{\frac{4}{3} \pi \mu m_{\mathrm{H}} \  r_{\mathrm{eff}}^3 }
\end{equation}

\begin{table*}
	\centering
	\caption{\centering{Core Observed Properties}}
	\label{Physical parameter}
	\begin{tabular}{ccccccccccc} 
		\hline
		\hline
Core &  RA(J2000) &  Dec(J2000) & Size$^{a}$ & Size & PA & $S_\mathrm{peak}$ & $S_\mathrm{core}$ & ${T_d}^{b}$ & ${M_\mathrm{{gas}}}$ & $n_{\mathrm{H_2}}$\\
 & (h:m:s) & (d:m:s) & ($^{\prime\prime}\times ^{\prime\prime}$) & (pc$\times$pc) & (deg) & (mJy/beam) & (mJy) & (K) & $(\mathrm{M_{\odot}})$ & $(\mathrm{cm}^{-3})$ \\
\hline
1 & 18:09:20.33	&  -20:14:55.6 &	 4.1$\times$3.0 & 0.06$\times$0.04 &  106 &  4.6 & 37.2 & 16.6 & 11.5 & $2.6\times10^6$\\
2 & 18:09:20.37	&  -20:15:05.0 &	 2.8$\times$2.0 & 0.04$\times$0.03 & 128 & 10.6 & 45.3& 16.6 & 14.0 & $1.0\times10^7$\\
3 & 18:09:20.76	&  -20:15:00.4 &	 3.5$\times$2.7 & 0.05$\times$0.04 & 13 & 8.3 & 55.6 & 16.6 & 17.2 & $5.9\times10^6$\\
\hline
 & 	&   &	  &  &   &   &  & $T_{\mathrm{H_2CO}}^{c}$ &  & \\
  & 	&   &	  &  &   &   &  & (K) &  & \\
\hline
1 & 	&   &	  &  &   &   &  & 16.6 & 11.5 & $2.6\times10^6$\\
2 & 	&   &	  &  &   &   &  & 83.0 & 2.1 & $1.5\times10^6$\\
3 & 	&   &	  &  &   &   &  & 67.7 & 3.2 & $1.1\times10^6$\\
\hline
 & 	&   &	  &  &   &   &  & $T_{\mathrm{typical}}^{d}$ &  & \\
   & 	&   &	  &  &   &   &  & (K) &  & \\
\hline
1 & 	&   &	  &  &   &   &  & 16.6 & 11.5 & $2.6\times10^6$\\
2 & 	&   &	  &  &   &   &  & 40 & 4.7 & $3.6\times10^6$\\
3 & 	&   &	  &  &   &   &  & 40 & 5.8 & $2.0\times10^6$\\
\hline
	\end{tabular}
	\begin{tablenotes}
        \footnotesize
        \item a: Core size deconvolved from synthesized beam. \\ b: Dust temperature of G10.21 adopted from \citet{Yuan_2017}. \\  c: Rotational temperature by fitting the $\mathrm{para-H_2 CO}$ lines, see more details in Section \ref{Th2co}.     \\ d: Typical dust temperature for protostars, see more details in Section \ref{Th2co}.

      \end{tablenotes}
\end{table*}

\begin{figure}[bt!]
    \centering
    \includegraphics[width=1.0\linewidth]{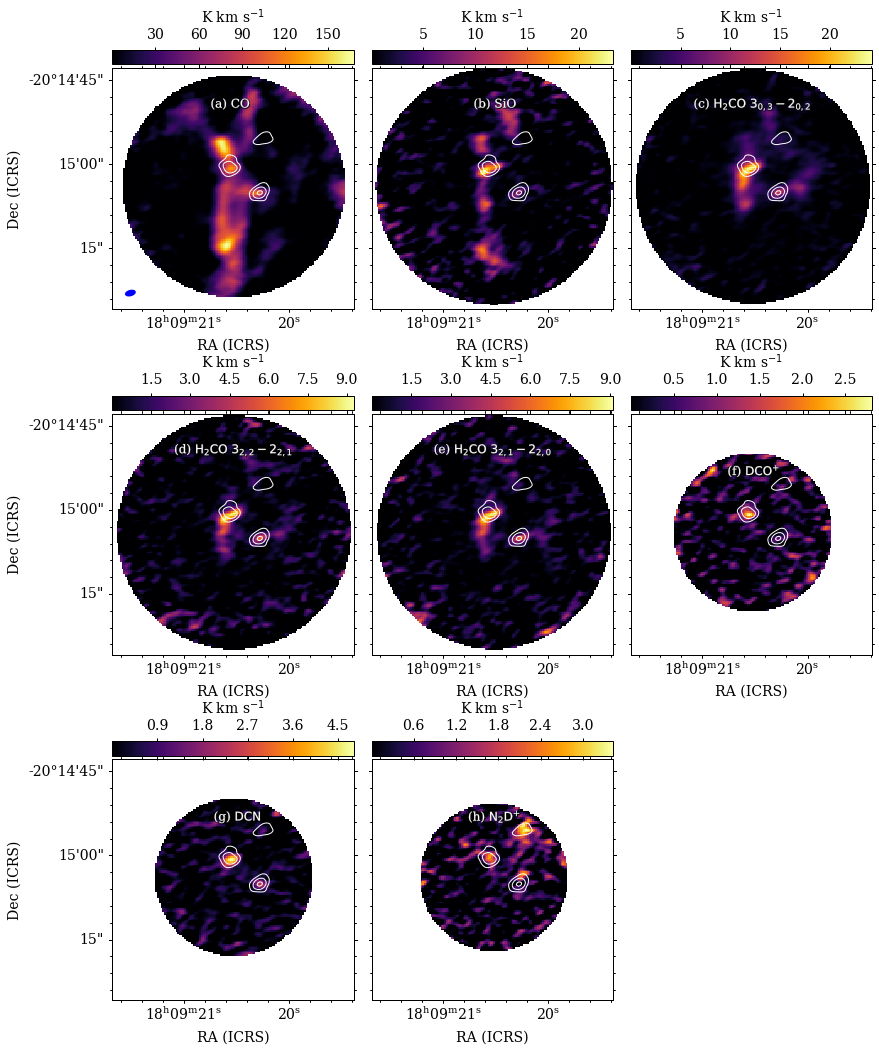}
    \caption{Moment-0 maps of different molecular lines after primary beam correction. We cut the image at the primary beam response of 0.2 for CO and SiO, and 0.5 for other lines, leading to little differences in field of view. The integrated velocity range is from -15 $\mathrm{km \ s^{-1}}$ to 30 $\mathrm{km \ s^{-1}}$ for CO and SiO, 4 $\mathrm{km \ s^{-1}}$ to 20 $\mathrm{km \ s^{-1}}$ for three para-$\mathrm{H_2 CO}$ lines, and from 8 $\mathrm{km \ s^{-1}}$ to 16 $\mathrm{km \ s^{-1}}$ for other lines. The white contours show continuum emission at levels of [6, 12, 24, 48]$\sigma$, where $\sigma$ equals to $5 \times 10^{-4} \, \mathrm{Jy/beam}$. The top-left blue ellipse in the bottom left marks synthesized beam.}
    \label{fig:line-M0}
\end{figure}

where $\mu$ ($\mu$ = 2.37) is the mean molecular weight per free particle, considering $\mathrm{H_2}$, He, and ignoring the number of heavier elements \citep{Kauffmann_2008},  $m_\mathrm{H}$ is the mass of a hydrogen atom, and $r_{\mathrm{eff}}=\frac{d}{2}\sqrt{\Theta_\mathrm{maj}\Theta_\mathrm{min}}$ is the equivalent radius of each core. The mean number density of cores is in range from $10^6$ to $10^7\ \mathrm{cm}^{-3}$ as listed in Table \ref{Physical parameter}. The mean number density of Core 2 and Core 3 is several times larger than Core 1.

The derived core mass ranges from $11 \ \mathrm{M_{\odot}}$ to $18 \ \mathrm{M_{\odot}}$. Here we discuss the uncertainties of derived mass. We can get the uncertainty of continuum flux and effective radius in \emph{imfit} function ($\sim 10 \%$). The typical uncertainties of gas-to-dust ratio and dust opacity are 23$\%$ and 28 $\%$ derived in \citet{Sanhueza_2017}. Using the online Parallax-Based Distance Calculator\footnote{http://bessel.vlbi-astrometry.org/bayesian}, the uncertainty in the distance is 7$\%$. The uncertainty of dust temperature is taken to be 10 $\%$. Taking all these things into account, we estimate a mass and number density uncertainty of $\sim 42\%$ and $\sim 56\%$. 


\subsection{Molecular Line Emission}
\par Different molecular lines trace different physical conditions, providing useful information about dense cores and their surrounding environments. The ALMA observations cover a total of $\sim$8 GHz bandwidth in four spectral windows, detecting many molecular lines in dense cores.  We show the spectra of three identified cores and discuss their chemical differential in Appendix \ref{sec:core_spectra}. The differential of line richness suggests an evolution path from Core 1 to Core 3.
\par Figure \ref{fig:line-M0} shows the integrated intensity map of molecular lines used in our analysis. Since CO, SiO and $\mathrm{H_2CO}$ trace more extended emission, deuterated species are good dense gas tracers. We cut the image at the primary beam response of 0.2 for CO, SiO, $\mathrm{H_2CO}$, and 0.5 for deuterated molecular lines, leading to little differences in field of view. The integrated velocity ranges from -15 $\mathrm{km \ s^{-1}}$ to 30 $\mathrm{km \ s^{-1}}$ for CO and SiO, 4 $\mathrm{km \ s^{-1}}$ to 20 $\mathrm{km \ s^{-1}}$ for three para-$\mathrm{H_2 CO}$ lines, and from 8 $\mathrm{km \ s^{-1}}$ to 16 $\mathrm{km \ s^{-1}}$ for three deuterated molecular lines ($\mathrm{DCN}$, $\mathrm{DCO^+}$, $\mathrm{N_2D^+}$). 
\par The spatial distribution of CO and SiO is mainly associated with outflow activities and less associated with continuum emissions. Note that SiO emission can also be emitted by accretion disks \citep{Maud_2018}. In our analysis, we exclude the possibility because of the large spatial scales ($> 10^4 \ \mathrm{AU}$). Previous studies suggest $\mathrm{H_2CO}$ lines can be used to trace not only the compact cores but also outflows (e.g., \citealt{tychoniec_2019,Beuther_2021}), which is consistent with our observations. The spatial distribution of three deuterated lines mainly agrees with the continuum emission, indicating that the three molecular lines are good dense gas tracers.

\subsection{Outflow Properties}
\label{outflow}
\begin{figure}[bt!]
    \centering
    \includegraphics[width=0.45\textwidth]{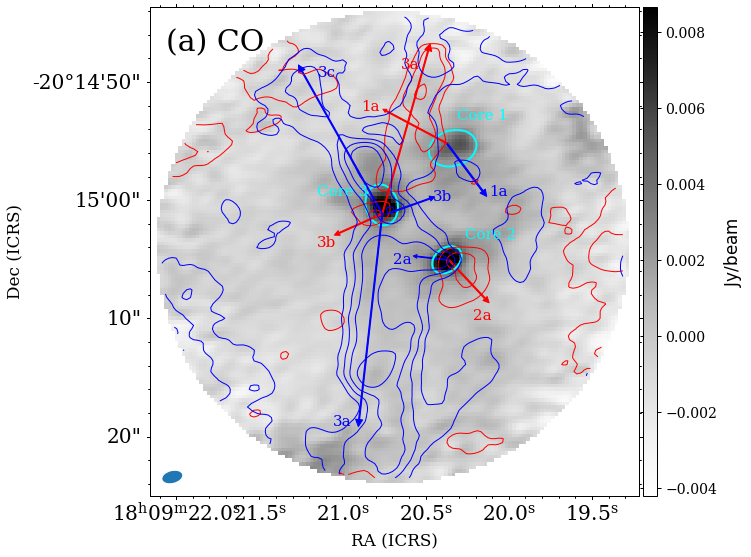}
    \hspace{0in}
    \includegraphics[width=0.45\textwidth]{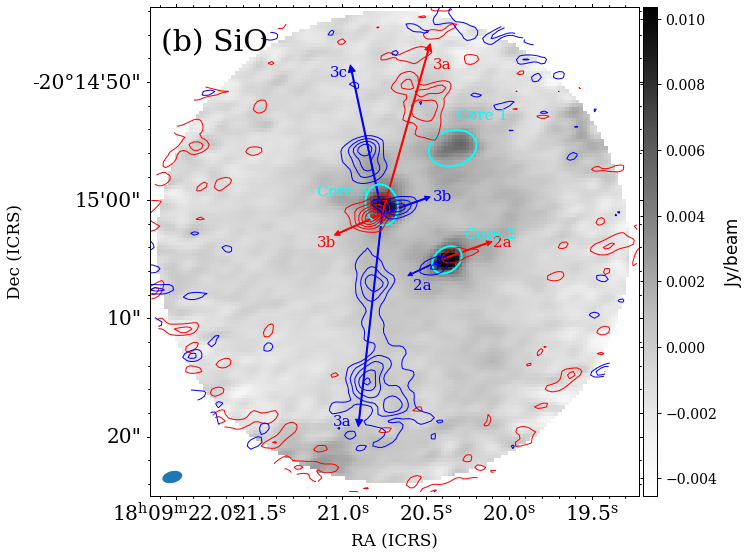}
    \caption{CO and SiO outflows of G10.21 after primary beam correction. The grayscale background image shows the ALMA 1.3 mm continuum emission. The blue ellipse in the bottom-left marks the beam area of the continuum. The cyan ellipse marks the position of the cores. \textbf{(a)}: CO outflows. The blue and red contours show the blueshifted and redshifted CO emission integrated over [-7, 8] $\mathrm{km \ s^{-1}}$ and [14, 29] $\mathrm{km \ s^{-1}}$, at levels of [3, 15, 30, 50, 100, 200]$\sigma$ and [6, 15, 30, 50, 100, 200]$\sigma$, where $\sigma$ equals to $2 \ \mathrm{K} \  \mathrm{km} \ \mathrm{s}^{-1}$.  \textbf{(b)}: SiO outflows. The blue and red contours show the blueshifted and redshifted SiO emission integrated over [-10, 10] $\mathrm{km \ s^{-1}}$ and [14, 34] $\mathrm{km \ s^{-1}}$, at levels of [3, 6, 9, 12, ...]$\sigma$, where $\sigma$ equals to $1 \ \mathrm{K} \ \mathrm{km} \ \mathrm{s}^{-1}$. The blue and red arrows mark the directions of outflows.}
    \label{fig:SiO_outflow}
\end{figure}

The outflow signatures of three compact cores are revealed by CO (2-1) and SiO (5-4) emission. To identify the outflows, we concentrate on blueshifted and redshifted CO and SiO emission relative to the systematic velocity of the sources. Figure \ref{fig:SiO_outflow} shows the velocity-integrated emission map and identified outflow lobes of CO (2-1) and SiO (5-4). The detailed process of outflow identification for the complex CO emission is shown in Appendix \ref{CO_outflow_identification}. A slight difference has been noted in the number and orientation of the identified lobes using these two tracers. Note that CO and SiO have different excitation temperatures and thus, a slight difference in the identified lobes is possible. We find explicit bipolar outflow activities towards Core 2 and Core 3, indicating the two cores are associated with ongoing star formation activities. We also find more than one group of outflows associated with Core 3, which implies that there may be multiple driving sources within this core. In addition, we find a possible weak CO outflow around Core 1 that is however lower than the velocity range we had defined for characterizing outflow emission. Also, it may be attributed to side lobe contamination or an extension of outflow lobe o3a. Therefore, the evolutionary phase of Core 1 is uncertain. 

\par Since the emission from CO lobes is tangled and it is difficult to distinguish them from each other, we use SiO data to estimate the outflow parameters only for those lobes that are identified in both tracers. For the red lobe of outflow 3a, we assume that the observed emission is associated with Core 3 rather than Core 1 because of two possible reasons. First, no blue lobes are identified for Core 1 in SiO and second, the intensity of the blue lobe of 3a is strong enough to make us consider that the red emission only corresponds to the red lobe of 3a. We calculate the SiO column density according to the equation from \citet{Mangum&Shirley_2015}. Assuming that the beam filling factor equals to 1, SiO emission is optically thin, the temperature of the background source is negligible, Rayleigh–Jeans approximation and local thermodynamic equilibrium (LTE) conditions can be applied, we can get the following equation:
\begin{equation}
     N_{\mathrm{tot}}={(\frac{3 k_B}{8 \pi^{3} \nu S \mu_d^{2} }) \  (\frac{Q_{\mathrm{rot}}}{g_{J} g_{K} g_{I}}) \exp({ \frac{E_{u}}{k_B T_{e x}}}) \int {T_\mathrm{B} d v}}
\end{equation}
where ${k_B}$ is the Boltzmann constant, $\nu$ is the rest frequency of the SiO (5-4) transition, $S$ is the line strength, $\mu_d$ is the permanent dipole moment of the molecule, $Q_\mathrm{rot}$ is the partition function of the molecule, $g_i$ are the degeneracies, $E_u$ is the energy of the upper energy level, $T_{ex}$ is the excitation temperature. Here we can simplify the equation:
\begin{equation}
\label{Column_density}
     N_{\mathrm{SiO}}\left(\mathrm{cm}^{-2}\right)=1.54 \times 10^{10}\left(T_{\mathrm{ex}}+0.347\right) \exp \left(\frac{31.26}{T_{\mathrm{ex}}}\right) \int T_{\mathrm{B}} d v
\end{equation}
In our calculation, we assume $T_\mathrm{ex} \approx T_\mathrm{kin,NH_3}$=18.5 K. Then we can derive the outflow parameters following the similar procedure reported in \citet{Wang_2011} and \citet{Tapas_2021}:
\begin{equation}
     M_{\mathrm{out}} =\frac{d^{2}}{\left[\frac{\mathrm{SiO}}{\mathrm H_{2}}\right]} \mu \mathrm{m_H} \int_{\Omega} N_{\mathrm{SiO}}\left(\Omega^{\prime}\right) d \Omega^{\prime} 
\end{equation}
\begin{equation}
     P_{\mathrm{out}} =M_{\mathrm{out}} v \times \frac{1} {\mathrm{cos} \ i}
\end{equation}
\begin{equation}
     E_{\mathrm{out}} =\frac{1}{2} M_{\mathrm{out}} v^{2} \times \frac{1} {\mathrm{cos}^2  i}
\end{equation}
\begin{equation}
     t_{\mathrm{dyn}} =\frac{l_{\mathrm{flow}}}{v_{\mathrm{Lobe}}} \times \frac{\mathrm{cos} \ i} {\mathrm{sin} \ i}
\end{equation}
\begin{equation}
     \dot{M}_{\mathrm{out}} =\frac{M_{\mathrm{out}}}{t_{\mathrm{dyn}}}\times \frac{\mathrm{sin} \ i} {\mathrm{cos} \ i}
\end{equation}
where $d$ is the kinetic distance to G10.21, which is taken to be 3.1 kpc \citep{Yuan_2017},  $\left[\frac{\mathrm{SiO}}{\mathrm{H_2}}\right]$ is the SiO abundance relative to ${\mathrm{H_2}}$. In our calculation, we set the value to be $1.8 \times 10^{-10}$, which is the average SiO abundance for infrared-quiet sources in \citet{Csengeri_2016}. Note that the SiO abundance may be enhanced in outflow regions and vary greatly in different regions, which may cause a few orders of magnitude difference from $10^{-12}$ to $10^{-8}$ (e.g., \citealt{Li_2019_a,Lu_2021}). The $v$ is the velocity of the outflow relative to the systematic velocity of driving source in the line of sight and the systematic velocity is determined by $\mathrm{N_2D^+}$ line fitting (11.5  $\mathrm{km \ s^{-1}}$ for Core 2 and 12.3 $\mathrm{km \ s^{-1}}$ for Core 3). ${l_\mathrm{flow}}$ is the maximum distance between the extent of outflow lobe and the central source projected to the plane of sky. In addition, ${v_\mathrm{lobe}}$ is the maximal velocity offset in the line of sight of the outflow lobe relative to the driving source and $i$ is the inclination angle between the outflow jet and the line of sight, which is set to be an average value of $57.3^{\circ}$ assuming all orientations are equally favorable (see \citealt{Bontemps_1996} for detailed calculation).

\begin{table}[!ht]
	\caption{\centering{Derived Outflow Parameters}}
	\centering
	\label{outflow parameter}
	\begin{tabular}{ccccccccc} 
		\hline
\multirow{2}{*}{Parameter$^a$} &  \multicolumn{2}{c}{o2a} &  \multicolumn{2}{c}{o3a} & \multicolumn{2}{c}{o3b} & \ o3c$^b$ \\
~ & Blue & Red & Blue & Red & Blue & Red & Blue   \\
\hline
$v \ (\mathrm{km \ s^{-1}}$) & [-11.9, 10] & [13.1, 24.9] & [-4.1, 9.0] & [13.1, 21.8] & [-0.1, 12.0] & [13.1, 25.9] & [-5.1, 10] \\

${M_\mathrm{out}} \ (\mathrm{M_\odot})$  & 0.58 & 0.38 & 9.31 & 3.81 & 1.09 & 2.93 & 2.79 \\

${P_\mathrm{out}} \ (\mathrm{M_\odot \ km \ s^{-1}})$  & 13.37 & 4.98 & 152.2 & 32.43 & 10.71 & 30.40 & 49.13 \\

${E_\mathrm{out}} \ (\mathrm{M_\odot \ km^2 \ s^{-2}})$  & 194.0 & 38.28 & 1450 & 171.7 & 73.76 & 220.2 & 509.0 \\

${l_\mathrm{flow}} \ (\mathrm{pc})$  & 0.04 & 0.05 & 0.31 & 0.17 & 0.04 & 0.05 & 0.12 \\

${t_\mathrm{dyn}} \ (\mathrm{10^3 \ yr})$  & 1.07 & 2.34 & 11.87 & 11.24 & 2.03 & 2.31 & 4.33 \\

${\dot M_\mathrm{out}} \ (\mathrm{10^{-4} \ \mathrm{M_\odot} \ yr^{-1}})$  & 5.36 & 1.62 & 7.84 & 3.39 & 5.38 & 12.67 & 6.43 \\


\hline
	\end{tabular}
	 \begin{tablenotes}
        \footnotesize
        \item a: All the values except $l_\mathrm{{flow}}$ have been corrected for inclination. \\
        b: No SiO red lobe is detected for outflow o3c.
        
      \end{tablenotes}
\end{table}

\par Table \ref{outflow parameter} lists the derived outflow parameters with the correction of inclination. Here we use a single excitation temperatures for all the outflows. In fact, the excitation temperature may be very different in different parts of outflow lobes  \citep[e.g.,][]{Green_2011}. We test the different excitation temperature in a range of 15-50 K and find that it would cause a maximum of $\sim 20 \%$ difference to our estimated parameters, which indicates the variations of excitation temperature only have a small effect on results. To check the validity of optically thin assumption, we derive the optical depth of SiO line using the RADEX\footnote{http://var.sron.nl/radex/radex.php} Non-LTE molecular radiative transfer online tool. Because the profile of SiO is not gaussian, we use the Moment-2 value of each outflow lobe as velocity dispersion. The regions of outflow lobes are away from the center of cores, the number density of the $\mathrm{H_2}$ gas would be much smaller than the average number density of each core, which is taken to be $10^4 \mathrm{cm^{-2}}$. We use our derived SiO column densities to calculate the optical depth of each lobe. The derived optical depths are smaller than 1 (range from 0.06 to 0.79), except for o3a, which indicates that the optically thin assumption of SiO is relatively reasonable. For outflow o3a, the two lobes are optically thick, resulting in underestimations for the derived parameters.
\par The outflow mass-loss rates in Table \ref{outflow parameter} are in orders of $\mathrm{10^{-4} \ \mathrm{M_\odot} \ yr^{-1}}$, comparable to the outflow mass-loss rates observed in several high-mass star forming regions (e.g., \citealt{Zhang_2005,Wang_2014,Liu_2017}), while they are more than three order magnitudes higher than those in low-mass star formation regions (e.g., \citealt{Bao_2014,Miranda_2020}). Assuming outflows are powered by accretion disks, we can infer the mass accretion rates according to the outflow force derived from SiO (5-4) \citep{Bontemps_1996}:
\begin{equation}
    \dot{M}_{\mathrm{acc}}=\frac{1}{f_{\mathrm{ent}}} \frac{\dot{M}_{\mathrm{acc}}}{\dot{M}_{w}} \frac{1}{V_{w}} \frac{\mathrm{P_{out}}}{\mathrm{t_{dyn}}}
\end{equation}
where ${f_{\mathrm{ent}}}$ is the entrainment effiiciency relating the SiO outflow source to the momentum flux of the wind at its source. The value of ${f_{\mathrm{ent}}}$ is typically in range of [0.1, 0.25] and we take it as 0.25 \citep{Liu_2017}. $\frac{\dot{M}_{{w}}}{\dot{M}_\mathrm{acc}}$ is the ratio of the wind/jet mass-loss rate to the mass accretion rate and the typical value is $\sim 0.1$ based on magneto-hydrodynamic models (e.g., \citealt{Konigl_2000,Cabrit_2009}). After assuming a typical jet/wind velocity of ${V_{w}} \sim 500 \ \mathrm{km \ s^{-1}}$, we can estimate the total accretion rates of o2a, o3a, o3b, o3c are about $5.7 \times 10^{-4}$, $6.0 \times 10^{-4}$, $7.2 \times 10^{-4}$, $8.9 \times 10^{-4}$ $\mathrm{ \mathrm{M_\odot} \ yr^{-1}}$ and the uncertainties of the estimated mass accretion/outflow rates are mainly caused by the uncertainties of the parameters discussed above. Compared to previous studies, the mass accretion rates here are comparable to some massive star formation models (e.g., \citealt{McKee_2003,Wang_2010}) and some other outflow studies in high-mass star forming regions (e.g., \citealt{Zhang_2005,Qiu_2009,Liu_2017,Lu_2018}).

The uncertainties of derived outflow parameters can come from many aspects such as the SiO abundance, outflow inclination angle, error in measurement, optically thin assumption, and adopted excitation temperature. Here we only consider the uncertainties caused by three main aspects. First, the typical uncertainty of SiO abundance is considered to be a factor of 10 ($\sim$ 1 dex), and the maximum value can be up to two order of magnitudes. Second, considering an angle range from 15 degrees to 75 degrees, the uncertainty caused by inclination angle will range from 2 to 5 due to the different formula forms about angles ($\sim$ 0.3-0.7 dex). Third, the error in measurement is considered to be 50\% ($\sim$ 0.2 dex), including uncertainties of flux and distance. Taking all these things into account, we estimate an uncertainty of 1.5-1.9 dex (a factor of 30-80) in the derived outflow and accretion parameters except for the dynamical timescale, and note that the uncertainty will be even larger in some regions. Because of the large uncertainties, we will not discuss these results further. The uncertainty of dynamical timescale only comes from the inclination angle and the error in measurement. We do not detect SiO outflow around Core 1 and the dynamical timescale of Core 2 and Core 3 is about $10^3$ and $10^4$ years. From the calculation of dynamical timescale, we derive an evolutionary picture from Core 1 to Core 3. Here we perform a Monte Carlo simulation to test the credibility of this conclusion. Assuming all orientations are equally favorable between 15 degrees to 75 degrees and adding an additional 50\% error of measurement, the probability that this conclusion still holds is $\sim$ 97\%. So the evolutionary sequence from Core 1 to Core 3 derived from dynamical timescale is relatively credible after considering the uncertainties.

\begin{figure}[t!]
    \centering
           \includegraphics[width=0.55\textwidth]{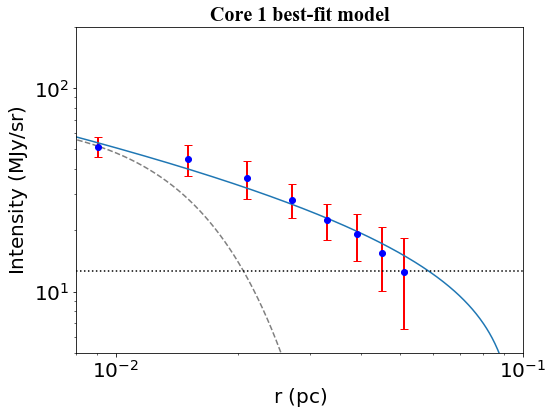}
            \label{fig:a}
            \includegraphics[width=0.55\textwidth]{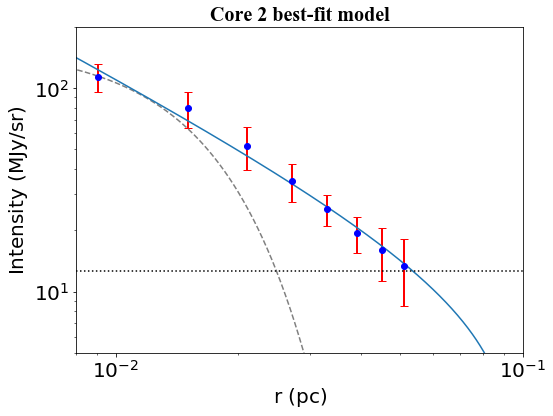}
            \label{fig:b}
            \includegraphics[width=0.55\textwidth]{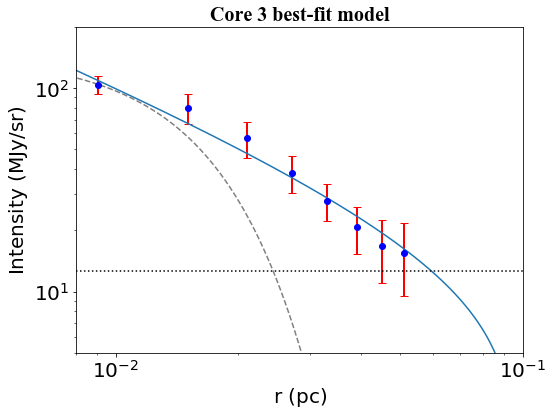}
    \caption{Best-fit intensity profiles of three cores. The blue line is the best-fit model and the blue points are the observational results. The grey dashed line represents the response of the beam and the black horizontal dotted line marks the $3 \sigma$ noise levels.}
    \label{density_profile}
\end{figure}

\subsection{Dense Core Structure}
\label{sec:core structure}
\par Using the combined ALMA 12m+7m continuum image, we can measure the intensity profiles of each core at 1.3 mm wavelength. We calculated the averaged intensity profiles in annuli with a width of $0.4^{\prime\prime}$ as a function of the projected distance to the core center (from $0.4^{\prime\prime}$ to $4^{\prime\prime}$). Because the maximum recoverable scale for 7 m data is $36^{\prime\prime}$, greatly larger than the core size, we can ignore the effect of missing flux. We smooth the continuum image to a circular beam of $1.72^{\prime \prime} \times 1.72^{\prime \prime}$ in advance to eliminate the effect of the elliptical beam shape. In order to estimate the density profile and derive the dynamical properties of each core, we assume that the density and temperature profiles both have power-law function forms: $\rho=\rho_{0}\left(r / r_{0}\right)^{-p}$ and $T=T_{0}\left(r / r_{0}\right)^{-q}$ within a maximum core radius $r_{\mathrm{max}}$. Under optically thin assumption, and Rayleigh–Jeans approximation, the intensity profiles can be derived analytically following the relation: $I_{\nu}(r) \propto r^{1-(p+q)}$ \citep[e.g.,][]{Beltran_2002}. The fitting results are listed in Table \ref{Core Structure}.
\par Because the optically thin assumptions may not be accurate for compact cores, we carry out radiative transfer analysis using RADMC-3D \citep{Radmc-3d} and compare the results with the observations to give a more accurate estimation of dense core structures and virial states. Here we briefly describe our models. We assume that both of the density and temperature profile have a power-law function form within a maximum core radius of 0.1 pc. If the core is internally heated, the temperature profile index q equals to 0.33 \citep{Scoville_1976}. However, for the edge part in our model, the external heating effect from the nearby HII regions can play an important role and we set the lowest dust temperature as 10 K in our models. We use the temperature derived in Section \ref{Th2co} to estimate temperature structures. For Core 1, we set $T_0=16.64$~K and $q=0$ because there is no clear evidence of internal heating.  For Core 2 and 3 with clear internal heating evidence, we set $q=0.33$ and $T_0$ equals to 79.5 and 58.8 K, respectively. There is no necessity to consider the index of dust opacity law because we only have dust continuum at a single wavelength. We assume dust opacities with the value of $\kappa=0.90 \mathrm{~cm}^{2} \mathrm{~g}^{-1}$ at 1.3 mm for $10^{6} \mathrm{~cm}^{-3}$ and range in different densities based on the OH5 models in \cite{OH94}. 
\par In summary, we need to fit two free parameters, the density in the reference radius ${r_0}$ ($\rho_0$) and the density profile index ($p$). The value of ${r_0}$ is arbitrary and we set ${r_0}=2000$ AU ($\sim 0.01$ pc)  in our calculations. 
\par The fitting procedure is similar to some previous studies (\citealt{Sanchez-Monge_2013,Palau_2014}). We make the sampling in two-dimensional parameter space and calculate the residuals in every step:
\begin{equation}
    \chi^{2} \equiv \sum_{i=1}^{n}\left[\frac{y_{i}^{\mathrm{obs}}-y_{i}^{\bmod }\left(\rho_{0}, p\right)}{\sigma_{i}}\right]^{2}
\end{equation}
The initial value is set to be $p=1.5 \pm 1.5, \ \rho_{0}=(1.0 \pm 1.0) \times 10^{-18} \mathrm{~g} \mathrm{~cm}^{-3}$. We run 1000 samples in one loop to find the best-fit values and reduce the step length by $20 \%$ in the next loop. The final best-fit parameters are selected after ten loops which consist of 10000 models. For the fitting with two free parameters, the uncertainty of each parameter can be estimated within the limit: $\Delta \chi^{2}=\chi^2-{\chi^2_{\mathrm{min}}} <2.3$. We show the best-fit results in Figure \ref{density_profile} and list the best-fit parameters in Table \ref{Core Structure}. Comparing the fitting results with analytical solution, we find that the analytical solution would overestimate the density profile index by $3 \%-20 \%$. Considering the possible CO outflow in Core 1, we also test the model that Core 1 is in protostellar phase. We set q=0.33 and maintain other parameters, and the derived density profile index and analytical results are 0.95 $\pm$ 0.26 and 1.34 $\pm$ 0.08. The density profile index is also underestimated and the proportion of underestimation reaches $\sim 40 \%$. The reason for the systematic difference is that the optical depth is assumed to be constant in the analytical solution. In fact, the optical depth is a function of the radius, decreasing as the radius increases. The density profile index ranges in 1.36$-$1.70, consistent with previous results of massive cores in high-mass star forming regions on $10^3-10^5$ au scales (e.g., \citealt{Wang_2011,Bulter_2012,Li_2019_b,Gieser_2021}). 

\begin{table}
	\centering
	\caption{\centering{Best-fit Parameters of Dense Core Structures}}
	\label{Core Structure}
	\begin{tabular}{m{1.2cm}m{1.5cm}m{4cm}m{2.8cm}m{1.5cm}m{1.5cm}m{2cm}ccccccc} 
		\hline
		\hline
{Core} & $q$ & \ \ \ \ \ \ \ \ \ \ {${\rho_0}^a$} &  \ \ \ \  \ \ $p^a$ & $\chi_{\mathrm{min}}^2$ &\  $\chi_r $ & \ \ \ \ \ \ ${p_1}^b$ \\

 & & \ \ \ \ \ \ $(\mathrm{g} \ \mathrm{cm}^{-3})$ & (RADMC) &  &  & (analytical)   \\
 
\hline 
1 & 0 &   $(1.7 \pm 0.5) \times 10^{-19}$ &$1.36 \pm 0.22$ &  1.18 & 0.44 & $1.67 \pm 0.08$  \\

\hline

2& 0.33 &   $(0.9 \pm 0.2) \times 10^{-19}$ & $1.70 \pm 0.24$ &  1.10 & 0.43 & $1.75 \pm 0.09$  \\

\hline

3& 0.33 &   $(1.0 \pm 0.2) \times 10^{-19}$ & $1.48 \pm 0.18$ &  2.48 & 0.64 & $1.63 \pm 0.10$  \\

\hline

\end{tabular}
\begin{tablenotes}
        \footnotesize
        \item a: free-parameters fitted in RADMC-3D. \\
        b: density profile index derived from analytically relation: $I_{\nu}(r) \propto r^{1-(p_1+q)}$.
      \end{tablenotes}
\end{table}

\begin{figure}[t!]
    \centering
           \includegraphics[width=0.6\textwidth]{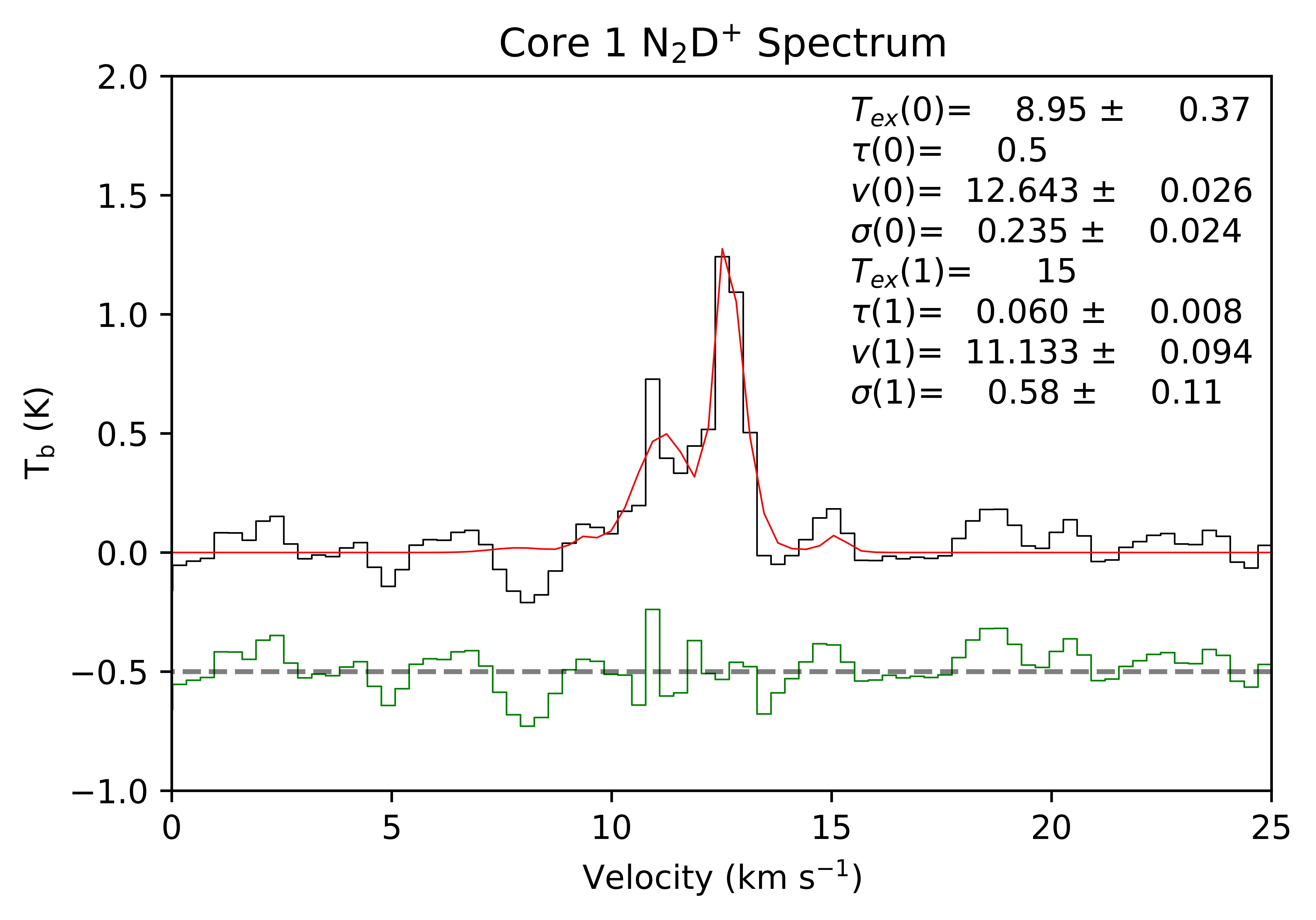}
            \label{fig:a1}
            \includegraphics[width=0.6\textwidth]{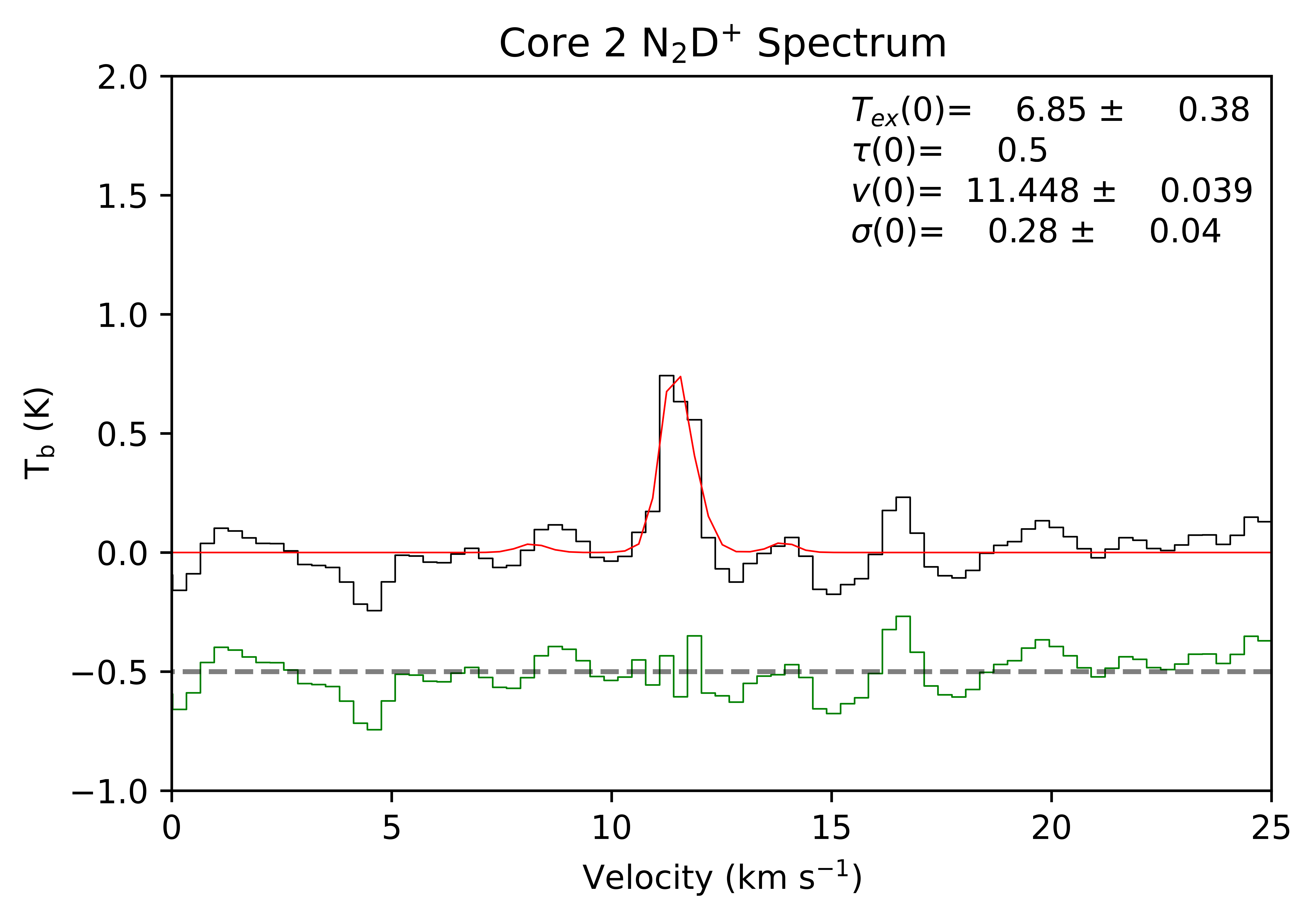}
            \label{fig:b1}
            \includegraphics[width=0.6\textwidth]{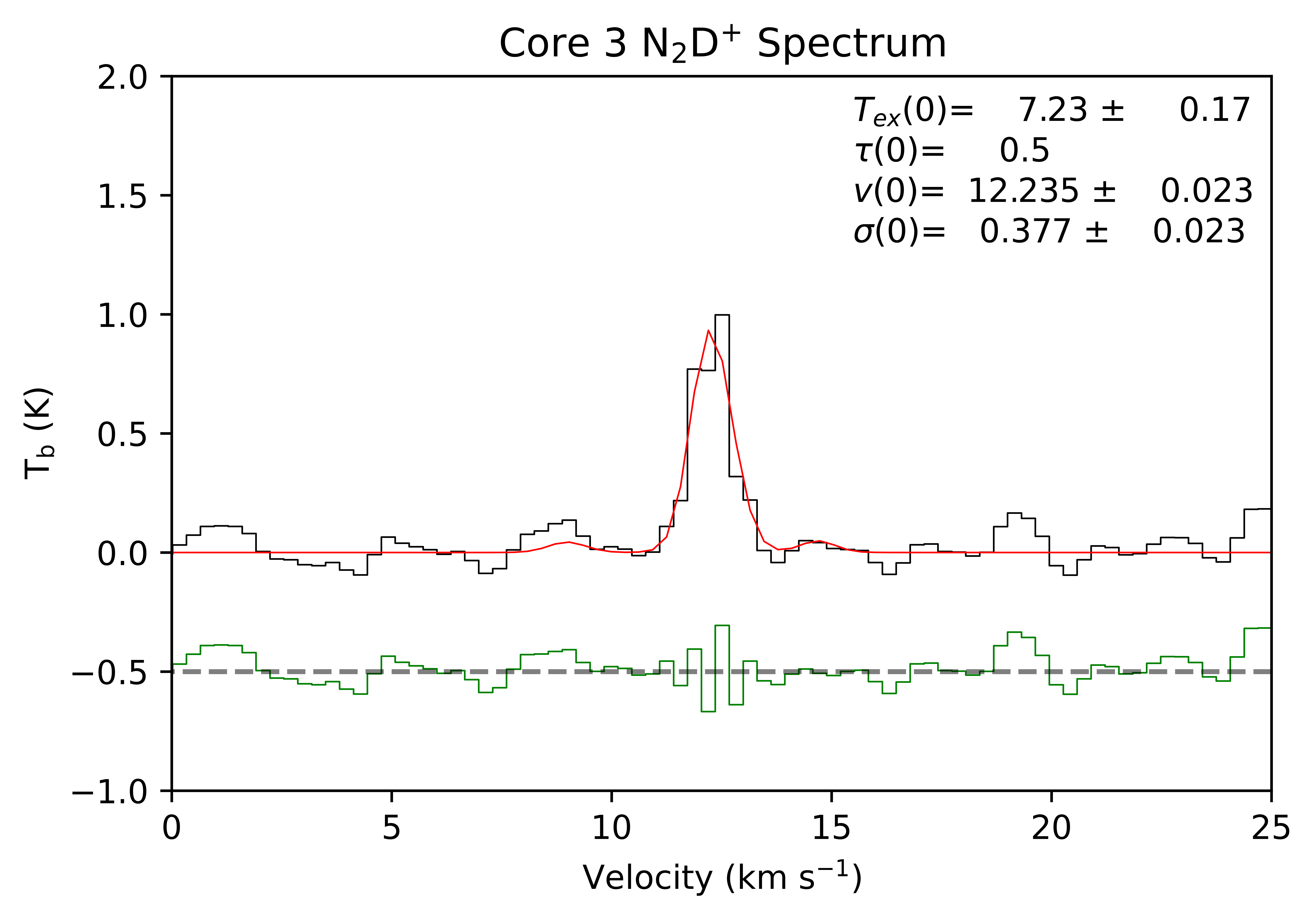}
    \caption{Core-averaged fitting results of $\mathrm{N_2D^{+}}$. The red line is the best-fit results and the green line is the fitting residual shifted by 0.5 K. The texts in the upper right are the best-fit parameters of hyperfine structures.}
    \label{n2dp}
\end{figure}

\subsection{Dynamical States of Cores}
\label{dynamical state}
Studies have shown that the abundance of $\mathrm{N_2D^+}$ remains high in the very early evolutionary stage of star formation (e.g., \citealt{Crapsi_2005,Kong_2015}) and $\mathrm{N_2D^+}$ is a good tracer to probe the dynamical states of prestellar/protostellar cores (e.g., \citealt{Kong_2017}). To derive the dynamical states and virial parameters of continuum cores, we need to fit $\mathrm{N_2D^+}$ spectra within three cores. We use Pyspeckit \citep{2011ascl.soft09001G} to fit hyperfine structures of $\mathrm{N_2D^+}$, deriving the centroid velocity and velocity dispersion. Under the optically thin assumption, the upper limit of optical depth is set to be 0.5 (the rationality of optically thin assumption can be seen in Section \ref{d_frac} for RADEX analysis). We detect $\mathrm{N_2D^+}$ in all three cores and the $\mathrm{N_2D^+}$ spectra of three cores are shown in Figure \ref{n2dp}. Core 2 and Core 3 have a single velocity component and Core 1 has two velocity components higher than $5\sigma$. We make $\mathrm{N_2D^+}$ channel maps in Appendix \ref{n2dp channel map} to visualize the data. As the Figure \ref{fig:n2dp_channel_map_pic} shows, all the velocity components are associated with the central cores and the two velocity components in Core 1 are not spatially resolved. For Core 1 with two velocity components, we estimate the gas mass of each velocity component assuming the continuum flux is proportional to the velocity-integrated intensity of the velocity component. Then we can derive the virial mass of each core (see details in \citealt{Bertoldi_1992,Li_2013}):
\begin{equation}
    M_{\mathrm{vir}}=\frac{5}{\alpha \beta} \frac{\sigma_{\text {tot }}^{2} r_\mathrm{eff}}{\mathrm{G}}
\end{equation}

where parameter $\alpha=(1-b / 3) /(1-2 b /5)$ is the correction to virial estimates for a power-law density profile $\rho \propto r^{-b}$ \citep{Maclaren_1988}, $\beta=\arcsin e / e$ is the geometry factor determined by eccentricity, $r_\mathrm{eff}$ is the effective radius of each core, $\mathrm{\sigma_{tot}}$ is the total velocity dispersion and can be derived as:
\begin{equation}
    \mathrm{\sigma_{tot}}=\sqrt{\left[\sigma_{\mathrm{obs}}^{2}-\Delta_{\mathrm{ch}}^{2} /(2 \sqrt{2 \ln 2})^{2}-\sigma_{\mathrm{th}, \mathrm{N_2D^+}}^{2}\right]+\sigma_{\mathrm{th},\mu \mathrm{m_H}}^{2}}
\end{equation}
where $\sigma_{\mathrm{th},\mu \mathrm{m_H}}=c_{\mathrm{s}}=\sqrt{\frac{k_{\mathrm{B}} T} {\mu m_{\mathrm{H}}}}$. Then we can derive the virial parameter $\alpha_{\mathrm{vir}}=M_{\mathrm{vir}} / M_{\mathrm{gas}}$ of each velocity component in compact sources. The uncertainties of dynamical parameters mainly come from three aspects: uncertainties of effective radius, uncertainties of derived velocity dispersion, and uncertainties of derived density profile index. The uncertainties of gas mass can be derived in Section \ref{sec:continuum}. Considering the possible difference of $\mathrm{N_2D^+}$ abundances in different velocity components of Core 1, we add an additional 20\% uncertainty in gas mass. As can be seen in Table \ref{Dynamical Parameter}, for the two cores with SiO outflow activities (Core 2 and Core 3), $\alpha_\mathrm{vir}$ is smaller than 0.5, which means that the two cores are gravitationally unstable. For the core at the earliest evolutionary stage (Core 1), the component with high centroid velocity is likely to undergo gravitational collapse and the component with low centroid velocity is in a critical state bound by gravity. If we consider the Core 1 at protostellar stage and apply $b$=0.95$\pm$0.26 into calculation, the derived virial parameters will be 1.9$\pm$1.1 and 0.3$\pm$0.1 for component 1 and component 2, respectively, which would not affect our conclusions. The different dynamical states may indicate unresolved structures. In addition, we also calculate the Mach number using $\mathcal{M}_s=\sqrt{3} \sigma_{\mathrm{nt}, \mathrm{N_2D}^{+}} / c_{\mathrm{s}}$. All the Mach numbers are higher than 1, which indicates general supersonic turbulence in this star-forming region.

\begin{table}
	\centering
	\caption{\centering{Core Dynamical Parameters}}
	\label{Dynamical Parameter}
	\begin{tabular}{m{0.8cm}m{2cm}m{1.5cm}m{2cm}m{2cm}m{1.5cm}m{1.5cm}m{1.5cm}cccccccc} 
		\hline
		\hline
{Core} &  {Component} &  \ \ \ \ $v$ & $\ \ \ \sigma_\mathrm{obs}$ & $\ \ \  \sigma_\mathrm{tot}$ & $ \  {M_\mathrm{vir}}$   & \  $ {M_\mathrm{gas}}$& \ \ $\alpha_\mathrm{vir}$  & $\mathcal{M}_s$  \\

 & & ($\mathrm{km \ s^{-1}}$) &  ($\mathrm{km \ s^{-1}}$) &  ($\mathrm{km \ s^{-1}}$) & \  $(\mathrm{M_{\odot}})$ & \  $(\mathrm{M_{\odot}})$ & & \\
 \hline 
\multirow{2}{*}{1} & \ \ \ \ \ \ 1 &  \ \ 11.1&  0.58$\pm$0.11 &  0.61$\pm$0.11 & 8.3$\pm$2.9 & 4.7$\pm$2.2 & 1.8$\pm$1.0 & 4.0$\pm$0.8 \\

  & \ \ \ \ \ \ 2 & \ \  12.6 &  0.24$\pm$0.02 & 0.30$\pm$0.02 & 2.0$\pm$0.3 & 6.8$\pm$3.1 & 0.3$\pm$0.1  & 2.2$\pm$0.1 \\

\hline

2& \ \ \ \ \ \ 1 & \ \  11.4 &  0.28$\pm$0.04 &  0.33$\pm$0.04 & 1.5$\pm$0.4 & 14.0$\pm$5.7 &0.1$\pm$0.05 & 1.7$\pm$0.3 \\

\hline

3& \ \ \ \ \ \ 1 & \ \  12.2 &  0.38$\pm$0.02 & 0.42$\pm$0.02 & 3.4$\pm$0.5 &17.2$\pm$7.1& 0.2$\pm$0.1 & 2.5$\pm$0.1 \\

\hline

\end{tabular}

\end{table}

\begin{figure}[bt!]
    \centering
    \includegraphics[width=0.43\textwidth]{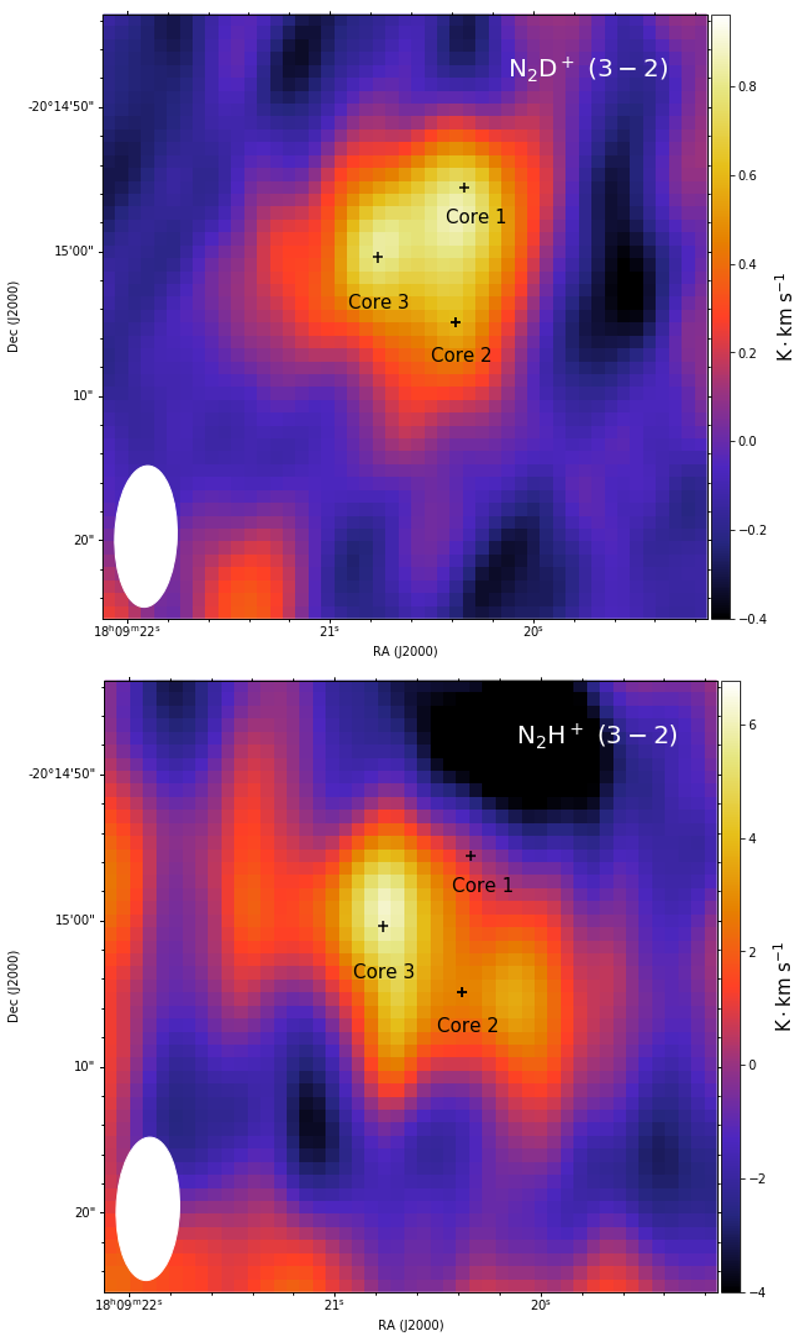}
    \hspace{0in}
    \includegraphics[width=0.47\textwidth]{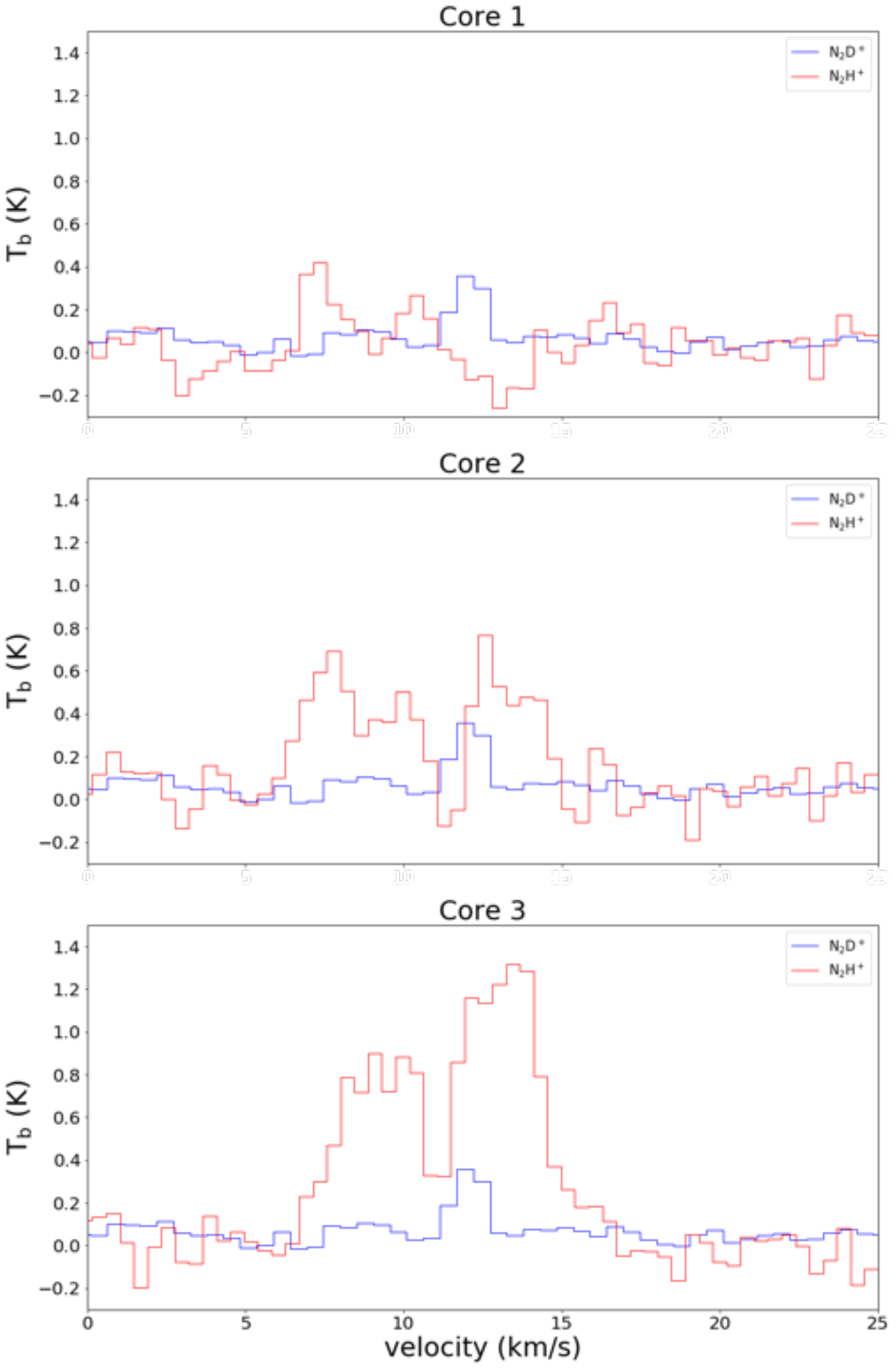}
    \caption{\textbf{Left}: Moment-0 maps of the $\mathrm{N_2D^+}$ and $\mathrm{N_2H^+}$. The integrated velocity ranges from 8 $\mathrm{km \ s^{-1}}$ to 15 $\mathrm{km \ s^{-1}}$. The white ellipse in the bottom left marks the SMA synthesized beam. The black crosses represent the position of the core center identified in ALMA data. \textbf{Right}: The spectra of the $\mathrm{N_2D^+}$ and $\mathrm{N_2H^+}$ extracted from the pixel where the core center is.}
    \label{fig:sma}
\end{figure}

\begin{table}
	\centering
	\caption{\centering{Deuterium Fraction of $\mathrm{N_2H^+}$}}
	\label{SMA_deuteration}
	\begin{tabular}{m{2cm}m{3cm}m{2.5cm}m{2.5cm}m{2cm}m{3cm}cccccc} 
		\hline
		\hline
Core &  $\sigma_\mathrm{T} \ (\mathrm{N_2D^+})$ &  \ \ ${N_{\mathrm{N_2D^+}}}$ & $\sigma_\mathrm{T} \ (\mathrm{N_2H^+})$  & \ \ \  ${N_{\mathrm{N_2H^+}}}$ & $\left[\mathrm{{N}_{2}D^+} / \mathrm{N_2H^+}\right]$\\
 & \ \ \ \ \  (K) & ($10^{11} \mathrm{~cm}^{-2}$) & \ \ \ \ \  (K)& ($10^{11} \mathrm{~cm}^{-2}$) &   \\
\hline
\ 1& \ \ \ \  0.036 & 5.7 (1.3)& \ \ \ \ 0.102 & \ 6.1 (3.5) &  \ \ 0.93 (0.62)\\
\ 2& \ \ \ \ 0.032 & 4.8 (1.0)& \ \ \ \ 0.088 & 24.9 (3.1) &  \ \ 0.19 (0.05) \\
\ 3& \ \ \ \ 0.032 & 7.0 (1.1)& \ \ \ \ 0.100 & 46.0 (3.5) &  \ \ 0.15 (0.03)  \\

\hline
	\end{tabular}
\end{table}

\subsection{Deuterium Fraction}
\label{d_frac}
Previous studies found that the deuterium fraction of $\mathrm{N_2H^+}$ decreases with time after protostellar stages (e.g., \citealt{Fontani_2011,Gerner_2015}). In this section we calculate the deuterium fraction of $\mathrm{N_2H^+}$ in three cores using SMA data. The SMA data covers both $\mathrm{N_2H^+}$ and $\mathrm{N_2D^+}$ lines, which provides good opportunities to compare our results with previous studies. Because the cores are barely resolved in these lines, we extract the spectra of $\mathrm{N_2H^+}$ and $\mathrm{N_2D^+}$ at the center of cores which are identified in Section \ref{sec:continuum}. Figure \ref{fig:sma} shows the integrated emission maps and the extracted spectra of three cores. We calculate the $\mathrm{N_2H^+}$ and $\mathrm{N_2D^+}$ column densities using equation \ref{Column_density} under the same assumptions with the excitation temperature equals to 15 K. In this calculation we don't distinguish the two velocity components in Core 1 as mentioned in Section \ref{dynamical state}. We also use the RADEX Non-LTE molecular radiative transfer tool to check the validity of optically thin assumption. The abundance of $\mathrm{N_2H^+}$ is much higher than $\mathrm{N_2D^+}$, using the derived column densities of $\mathrm{N_2H^+}$, we find the optical depth of each hyperfine component ranges from $10^{-4}$ to $5\times 10^{-2}$, indicating the optically thin assumption is reasonable for all three cores. To estimate the uncertainties of column densities, we use Monte Carlo method by adding to each pixel a gaussian noise centered at 0 K, with a standard deviation of $\sigma_\mathrm{T}$. Note that we mask the data points in Core 1 where $\mathrm{T_b}<0$ K because of the effect of side lobes. We repeat the calculations for 1000 times and take the standard deviation as the uncertainties of column densities. We list the final results in Table \ref{SMA_deuteration}. Although a low signal-to-noise ratio of $\mathrm{N_2H^+}$ cause an unreliable estimate of uncertainty for Core 1, the decreasing trend of deuterium fraction is clearly evident in the spectra. This indicates an early to late evolutionary sequence from Core 1 to Core 3.
\

\begin{figure}[t!]
    \centering
           \includegraphics[width=0.8\textwidth]{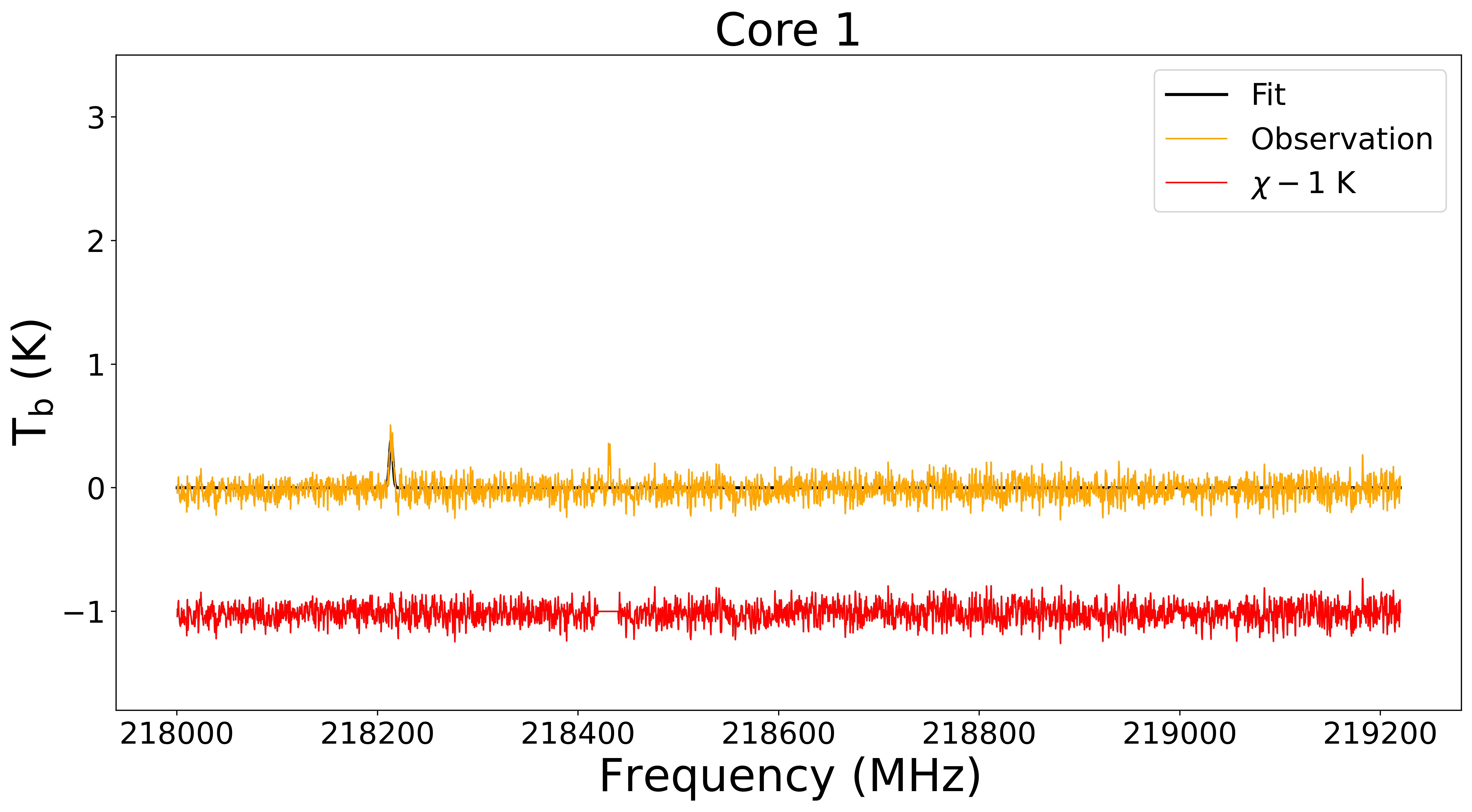}
            \label{fig:a2}
            \includegraphics[width=0.8\textwidth]{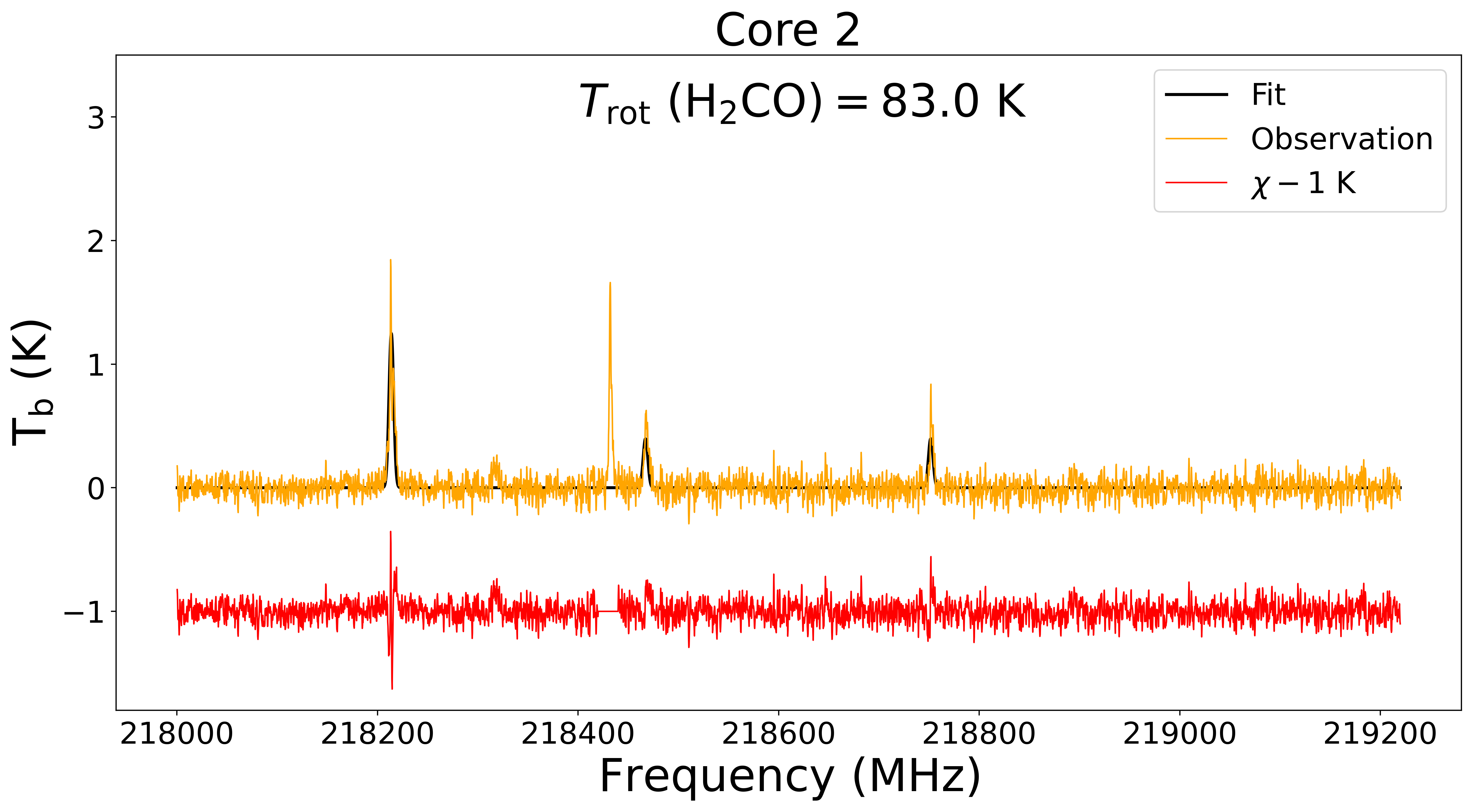}
            \label{fig:b2}
            \includegraphics[width=0.8\textwidth]{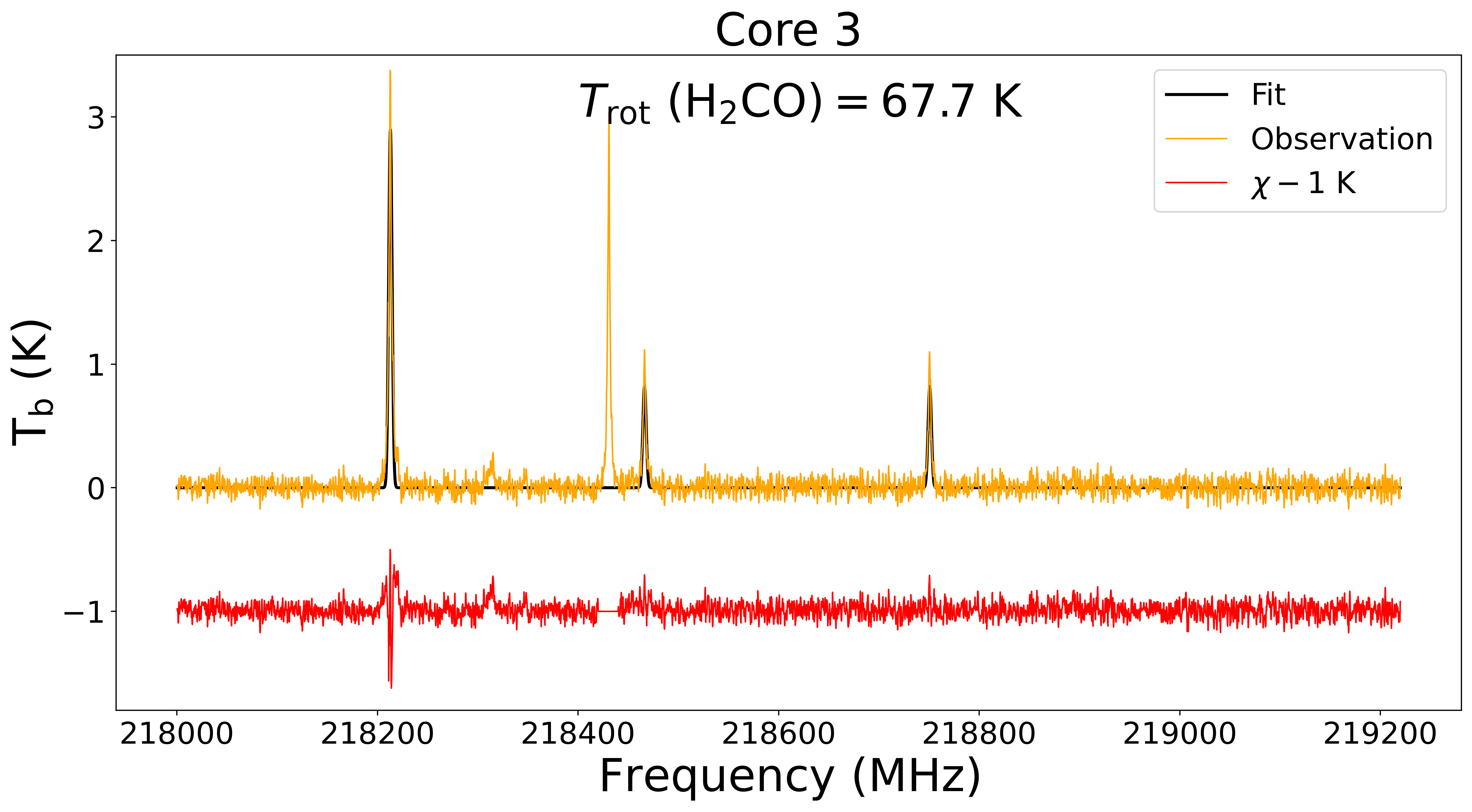}
    \caption{Core-averaged fitting results of $\mathrm{para-H_2 CO}$ lines. The black line is the best-fit results and red line marks the residual. The line around 218.43 GHz is one of $\mathrm{CH_3OH}$ lines which is not covered in this fit. Here we mask the residual of this line in the figure.}
    \label{H2CO}
\end{figure}

\section{Discussion}
\label{sec:discussion}

\subsection{Potential Bias from Temperature Estimation}
\label{Th2co}

\par In our analysis, we apply a uniform temperature to each core using the dust temperature of G10.21. However, we find SiO outflow activities in Core 2 and Core 3, which indicates that the three cores in G10.21 might be at different evolutionary stages. It is a crude assumption to consider a single dust temperature for all the cores and here we would like to give a correction to the temperature. Besides $\mathrm{NH_3}$, $\mathrm{H_2 CO}$ lines can also be used as a thermometer in dense molecular clouds \citep{Mangum_1993}. The $\mathrm{H_2 CO}$ lines around 218 GHz are easily measured and widely used in high-resolution interferometric observations \citep[e.g.,][]{Lu_2017,Beuther_2021}. Here we use the eXtended CASA Line Analysis Software Suite \citep[XCLASS,][]{Moller_2017} to fit the rotational temperature of $\mathrm{para-H_2 CO}$ lines. We fit the average spectra of each core and use the $\mathrm{T_{rot}} \ (\mathrm{H_2 CO}$) to replace the dust temperature. The line fitting is shown in Figure \ref{H2CO}. There is no clear detection of $\mathrm{para-H_2 CO}$ lines in Core 1, which indicates that the dust temperature around Core 1 is relatively low. The dust temperature applied in Core 1 is reasonable and we only apply the temperature correction in Core 2 and Core 3. However, the $\mathrm{para-H_2 CO}$ lines are easily affected by outflow activities \citep[e.g.,][]{Gomez_2013} and the outflow lobes are very close to the core center, which suggests that we may overestimate the dust temperature using the above correction. Here we also apply a temperature of 40 K (a typical temperature for protostars) towards the two cores and the results after temperature correction are listed in Table \ref{Physical parameter}. If we consider Core 1 at the protostellar stage and apply the temperature of 40 K to Core 1, the mass and density of Core 1 will be 3.9 $\mathrm{M_{\odot}}$ and $8.7 \times 10^5 \ \mathrm{cm^{-3}}$, respectively, making the mean number density of Core 1 several times smaller than Core 2 and Core 3. The uncertainties of derived parameters are similar to the previous discussion in Section \ref{sec:continuum}.

\begin{figure}[bt!]
    \centering
    \includegraphics[width=0.58\textwidth]{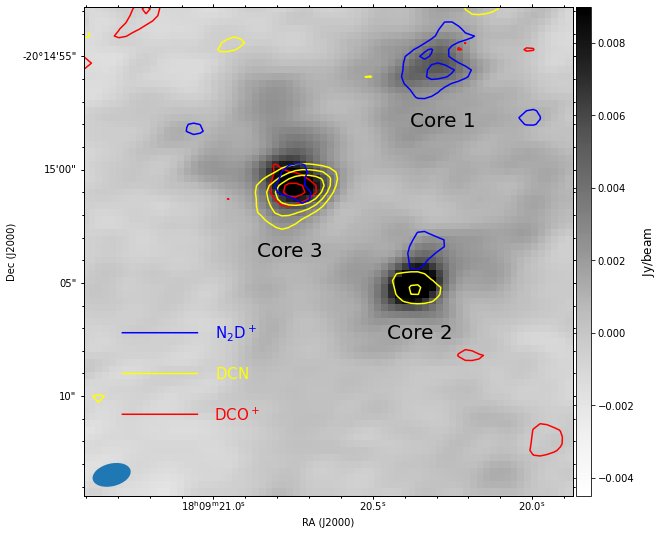}
    \hspace{0in}
    \includegraphics[width=0.32\textwidth]{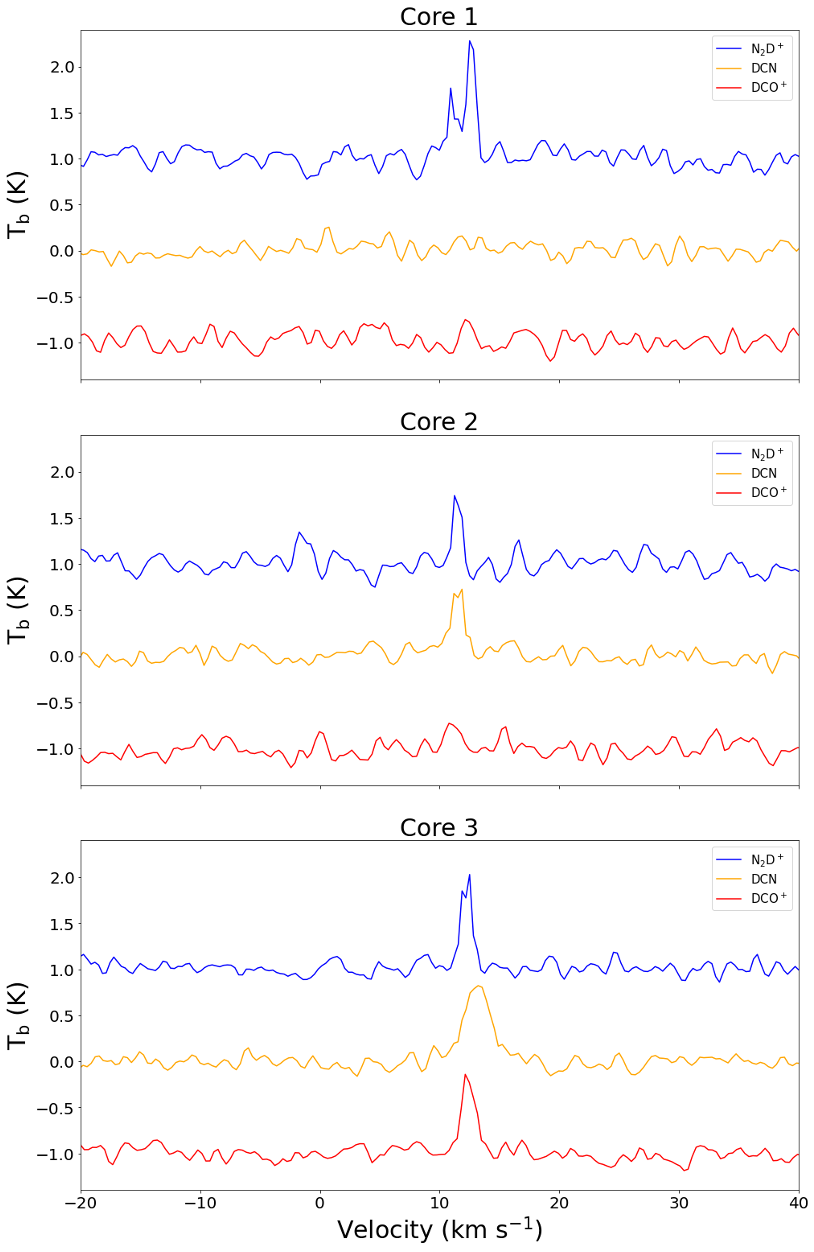}
    \caption{\textbf{Left}: Moment-0 maps of the three deuterated molecules after primary beam correction overlaid with 1.3 mm continuum map. The blue, yellow, red contours represent the emission of $\mathrm{N_2D^+}$, $\mathrm{DCN}$, $\mathrm{DCO^+}$, at levels of [3, 5, 7, 9,...]$\sigma$, where $\sigma$ equal to $0.62, \ 0.45, \ 0.47 \ \mathrm{K} \ \mathrm{km} \ \mathrm{s}^{-1}$, respectively.   \textbf{Right}: Core-averaged spectra of the three deuterated molecules within the three compact cores.}
    \label{fig:deuterated molecules}
\end{figure}

\par From Table \ref{Physical parameter} we can see that the mass of two protostars becomes relatively low after applying the temperature correction. Though we consider the overestimation from the fitting of $\mathrm{para-H_2 CO}$ lines, there are no more massive protostars in G10.21. The results may indicate that the core at the earliest stage contains the maximal mass, which is in contrast with the competitive accretion model \citep{Bonnell_2001}. However, we don’t know what percentage of the gas can eventually be fed into the collapse of the core. The evolution path of Core 1 still needs to be further explored. The density of dense cores at different evolutionary stages is comparable after applying the temperature correction. \citet{Kong_2021} investigate density change during high-mass star formation and find no difference when comparing the populations at different evolutionary stages, consistent with our results in Table \ref{Physical parameter}.

\begin{figure}[bt!]
    \centering
    \includegraphics[width=0.85\linewidth]{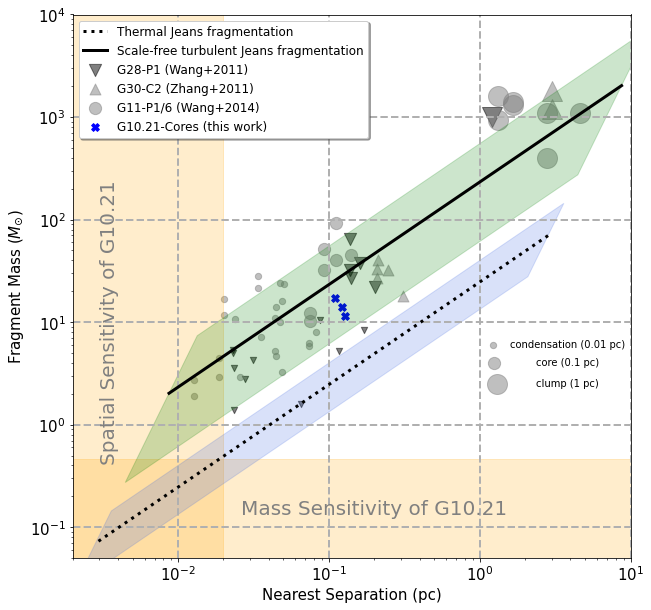}
    \caption{The relation between fragment mass and nearest separation distance. The grey downwards triangles mark the data of G28-P1 \citep{Wang_2011}. The grey upwards triangles represent the data of G30-C2 \citep{Zhang_2011}. The grey circles show the data of G11-P1 and G11-P6 \citep{Wang_2014}. The blue crosses and orange crosses mark the data in our work. The orange shaded regions show the sensitivity and resolution limit of the our ALMA observations. The dotted line shows thermal Jeans fragmentation with $T=18.5$ K and $n=[10^2,10^8]$ cm$^{-3}$, and the blue shaded region corresponds to the same density range but with $T=[10,30]$ K. The solid line shows a scale-free turbulent Jeans fragmentation with effective temperature $T_\mathrm{eff}$ of 175 K (total velocity dispersion $\sigma=0.78 \ \mathrm{km \ s^{-1}}$) and the same density range. The green shaded region corresponds to the same density range but with $T_\mathrm{eff}=[46,413]$ K (i.e., $\sigma=[0.4,1.2] \ \mathrm{km \ s^{-1}}$). The sizes indicate the physical scales of grey data points: the smallest are condensations ($\sim0.01$ pc), the middle are cores($\sim0.1$ pc), and the largest are clumps($\sim1$ pc). This figure shows clearly that the fragmentation in G10.21 are likely dominated by turbulence over thermal pressure at core scales under the sensitivity of our observations.}
    \label{fragment_mass}
\end{figure}

\subsection{Gas Fragmentation}
\par Fragmentation exists at different scales from giant molecular clouds to small gas clumps. In Section \ref{sec:continuum}, we find clear fragmentation in G10.21 and we will discuss how this source fragments into dense cores that have the potential to form high-mass stars.
\par If the fragmentation is dominated by thermal Jeans instability \citep{Jeans_1902}, the separation between the cores and the mass of cores will be comparable to the Jeans length and Jeans mass of G10.21. We calculate the Jeans length and Jeans mass following \cite{Wang_2014}:
\begin{equation}
\label{Jeans length}
    \lambda_{\mathrm{J}}=c_{s}\left(\frac{\pi}{G \rho}\right)^{1 / 2}=0.066 \  \mathrm{pc}\left(\frac{T}{10 \mathrm{~K}}\right)^{1 / 2}\left(\frac{n}{10^{5} \mathrm{~cm}^{-3}}\right)^{-1 / 2}
\end{equation}

\begin{equation}
\label{Jeans mass}
    M_{\mathrm{J}}=\frac{\pi^{5 / 2} c_{s}^{3}}{6 \sqrt{G^{3} \rho}}=0.877 \  \mathrm{M}_{\odot}\left(\frac{T}{10 \mathrm{~K}}\right)^{3 / 2}\left(\frac{n}{10^{5} \mathrm{~cm}^{-3}}\right)^{-1 / 2}
\end{equation}
where the temperature $T$ is the kinetic temperature derived from ammonia in Section \ref{sec:continuum} and density $n$ is the average density of G10.21 \citep{Yuan_2017}. The Jeans length and Jeans mass of G10.21 is about 0.04 pc and 0.97 $\mathrm{M_\odot}$. In our sample, the initial fragmentation leads to three massive cores with masses from 11 to 18 $\mathrm{M_\odot}$ with an average separation of 0.12 pc in the plane of sky. Considering the projection effects, the mean separation would be even larger than 0.12 pc. The large difference indicates that the fragmentation is not only controlled by gravity and thermal pressure, which is consistent with the conclusion derived in \citet{Zhang_2021}.
\par If we replace the $c_s$ by the total velocity dispersion of $\mathrm{NH_3}$ measured in \citet{Wienen_2012} ($\sigma=0.78 \ \mathrm{km \ s^{-1}}$), we can calculate the effective Jeans length and Jeans mass with the support of thermal pressure and turbulence. To investigate the role of turbulence in gas fragmentation, we use the similar method from \cite{Wang_2014} to visualize the relation between fragment mass and nearest separation distance. The results are shown in Figure \ref{fragment_mass}. This figure includes data points from previous relevant works at different spatial scales that can be compared with our measurements. The shaded blue region represents the thermal Jeans fragmentation domain where the adopted temperature and density are in ranges $T=[10,30]$ K and $n=[10^2,10^8]$\,cm$^{-3}$. The shaded green region represents the turbulent Jeans fragmentation domain with the same density range and effective temperature range $T_\mathrm{eff}=[46,413]$ K (i.e. total velocity dispersion $\sigma=[0.4,1.2] \ \mathrm{km \ s^{-1}} $). From Figure \ref{fragment_mass} we can conclude that the fragmentation in G10.21 is likely dominated by turbulence over thermal pressure at core scales within the sensitivity bounds of our observations, which is consistent with several previous works (e.g., \citealt{Zhang_2009,Pillai_2011,Wang_2011,Wang_2014}). Recent high-resolution observations down to the smallest accessible spatial scales find high-level fragmentations, more consistent with thermal Jeans fragmentation of dense cores (e.g., \citealt{Palau_2018,Beuther_2019,Tang_2021}). We cannot exclude the possibility that these cores may harbor further fragmentations not resolved, especially the SiO outflow image and different velocity components are consistent with the possibility. Further high resolution observations are needed to investigate the fragmentation mechanism at smaller scales.

\subsection{Deuterated Molecules as Chimecal Clocks}
\label{deuterated molecules}
\par Deuterated molecules are sensitive to temperature, which can provide useful information to probe the initial conditions (e.g., \citealt{Bacmann_2003,Pillai_2007,Li_2021}). Because of the different formation mechanisms, different deuterated species trace sources in different evolutionary stages. For example, the abundance of $\mathrm{N_2D^+}$ will decrease after the onset of star formation (e.g., \citealt{Caselli_2002,Fontani_2011}), while $\mathrm{DCO^+}$ and $\mathrm{DCN}$ are more likely to be detected in warm environments \citep{Gerner_2015}. The combination of the three deuterated molecules can be treated as chemical clocks, which has been proven by recent observations (e.g., \citealt{Morii_2021,Sakai_2021}). 
\par Figure \ref{fig:deuterated molecules} shows the distribution of the three deuterated molecules overlaid with the 1.3 mm continuum map and the spectra of the three cores. Core 1 shows strong $\mathrm{N_2D^+}$ emission but no $\mathrm{DCO^+}$ and $\mathrm{DCN}$ emission, which is considered at the earliest evolutionary stage in star formation. Core 2 exhibits weak $\mathrm{N_2D^+}$ and $\mathrm{DCO^+}$ emission, which indicates the core has just ignited star formation activities for a short period. Meanwhile, we can see clear $\mathrm{DCN}$ and $\mathrm{DCO^+}$ emission and weak $\mathrm{N_2D^+}$ emission within Core 3, which suggests Core 3 is at the latest evolutionary stage of star formation among the three cores. Here we estimate an evolutionary sequence from Core 1 to Core 3, consistent with the results derived from outflow activities. We don't give a quantitive correlation between the ratio of deuterated molecules and the dynamical timescale due to insufficient sample size. Detailed studies of the relation need to be investigated in a larger sample.

\subsection{The Potential for High-mass Star Formation}
\par G10.21 is considered as a high-mass starless core candidate, which is at the earliest stage of high-mass star formation. In this section we will discuss whether high-mass stars can form  in this source.
\par  According to \cite{Kauffmann_2010}, the region which has the potential to form high-mass stars should follow the relation: $m(r) \geqslant 580 \ \mathrm{M_{\odot}}(r / \mathrm{pc})^{1.33} $, after rescaling the dust opacities without the factor of 1.5 reduction (see the discussion in \citealt{Dunham_2011}). Applying the equivalent radius r=0.13 pc to the equation, the derived mass threshold is 38 $\mathrm{M_{\odot}}$. The mass surface density is another widely used parameter to estimate the potential of high-mass star formation. \cite{Urquhart_2013} suggest an empirical threshold for high-mass star formation of $0.05 \ \mathrm{g \ cm}^{-2}$. The mass and surface density of G10.21 are estimated to be $ 314 \ \mathrm{M_\odot}$ and $1.28 \ \mathrm{g \ cm}^{-2}$, greatly exceeding the above thresholds, which indicates that G10.21 has enough potential to form high-mass stars.


\par Since high-mass stars could form in G10.21, we can estimate the possible maximum stellar mass using the properties of G10.21. Based on the empirical relation mentioned in \cite{Larson_2003}: $\left(\frac{m_{\mathrm{max}}}{\mathrm{M}_{\odot}}\right)=1.2\left(\frac{M_{\text {cluster }}}{\mathrm{M_{\odot}}}\right)^{0.45}$, assuming a $30\%$ star formation efficiency, G10.21 could form a stellar cluster with a total stellar mass of $94 \ \mathrm{M_{\odot}}$. The derived maximum stellar mass is $9.3 \ \mathrm{M_{\odot}}$.  \cite{Sanhueza_2017} suggests another relation to estimate the maximum stellar mass using the the IMF from \cite{Kroupa_2001}: 
\begin{equation}
m_{\max }=\left(\frac{0.3}{\epsilon_{\text {sfe}}} \frac{17.3}{M_{\text {clump }}}+1.5 \times 10^{-3}\right)^{-0.77} \ \mathrm{M}_{\odot}
\end{equation}
Using the same $30\%$ star formation efficiency, a high-mass star with the mass of 9.1 $\mathrm{M_{\odot}}$ could form. 
 In Section \ref{sec:continuum}, the derived mass of identified cores ranges from 11.5-17.2 $\mathrm{M_{\odot}}$, slightly larger than the derived maximum stellar mass. The comparison indicates further fragmentation at smaller scales, which is consistent with our results (multiple outflows in Section \ref{outflow} and velocity components in Section \ref{dynamical state}).

\par Here we propose a possible evolutionary picture of G10.21. G10.21 is at a very early evolutionary stage of high-mass star formation due to its short dynamical timescale. It fragments into three dense cores, among them Core 3 starts the star-forming activities first and Core 1 is at the earliest evolutionary stage. No high-mass prestellar core is found in G10.21. Based on our results, we predict the three dense cores would fragment into more gas condensations at smaller scales. High-mass stars will eventually form in G10.21 upon the completion of gas accretion.

\section{Conclusion}
\label{sec:conclusion}
\par We study the fragmentation, core properties and chemical evolution towards a high-mass prestellar core candidate G10.21, using ALMA and SMA observations. Our main findings are as follows:
\par (1) We found three continuum compact sources in G10.21, with masses ranging from 11 $\mathrm{M_{\odot}}$ to 18 $\mathrm{M_{\odot}}$ at a uniform dust temperature of 16.6 K. We find a coherent evolutionary sequence from Core 1 to Core 3, based on line richness, outflow properties, deuterated molecules distribution, and deuterium fraction of $\mathrm{N_2H^{+}}$. No high-mass prestellar core is found in this source. This suggests a dynamical star formation where cores grow in mass over time.

\par (2) Several outflows are identified in SiO (5-4) and CO (2-1) lines. We derive the outflow parameters of lobes that are identified in both tracers, consistent with the previous work of other high-mass star forming regions. The dynamical timescale of Core 2 and Core 3 is roughly $10^3$ and $10^4$ yr using SiO outflows and we derive an evolution picture from Core 1 to Core 3.

\par (3) We derive the core structures using radiative transfer tools and compare the density profile index with it derived from the typical analytical method. The results are comparable, while the analytical method may overestimate the index by $3 \%-20 \%$. We also derive the virial parameters using the above index, finding different virial status of different cores. In addition, all the Mach numbers are higher than 2, suggesting general supersonic turbulence in G10.21. Turbulence is considered to play important roles in fragmentation at core scales in G10.21 within the sensitivity bounds of our observations.

\par (4) We derive the deuterium fraction of $\mathrm{N_2H^+}$ using SMA data. The deuterium fraction of $\mathrm{N_2H^+}$ decreases with the evolution of cores, which is consistent with previous works. We also derive similar conclusions through the spatial distributions of three deuterated molecules observed by ALMA. The combination of three deuterated molecules and outflow activities can be treated as chemical clocks, which need to be investigated quantitively in a larger sample in the future.


\section*{Acknowledgements}
We are grateful to an anonymous referee for the constructive
comments that helped us improve this paper. We acknowledge support from the China Manned Space Project (CMS-CSST-2021-A09, CMS-CSST-2021-B06), 
the National Key Research and Development Program of China (2017YFA0402702, 2019YFA0405100), the National Science Foundation of China (11973013, 11721303), and the High-Performance Computing Platform of Peking University. TGSP gratefully acknowledges support by the National Science Foundation under grant No. AST-2009842. TB acknowledges support from S. N. Bose National Centre for Basic Sciences, under the Department of Science and Technology, Government of India. This research has made use of the NASA/IPAC Infrared Science Archive, which is funded by the National Aeronautics and Space Administration and operated by the California Institute of Technology. This paper also makes use of the following ALMA data: ADS/JAO.ALMA$\#$2016.1.01346.S. ALMA is a partnership of ESO (representing its member states), NSF (USA) and NINS (Japan), together with NRC (Canada), MOST and ASIAA (Taiwan), and KASI (Republic of Korea), in cooperation with the Republic of Chile. The Joint ALMA Observatory is operated by ESO, AUI/NRAO and NAOJ. 

\facilities{ALMA, SMA, Spitzer, Herschel, APEX}

\begin{figure}[t!]
    \centering
           \includegraphics[width=1\textwidth]{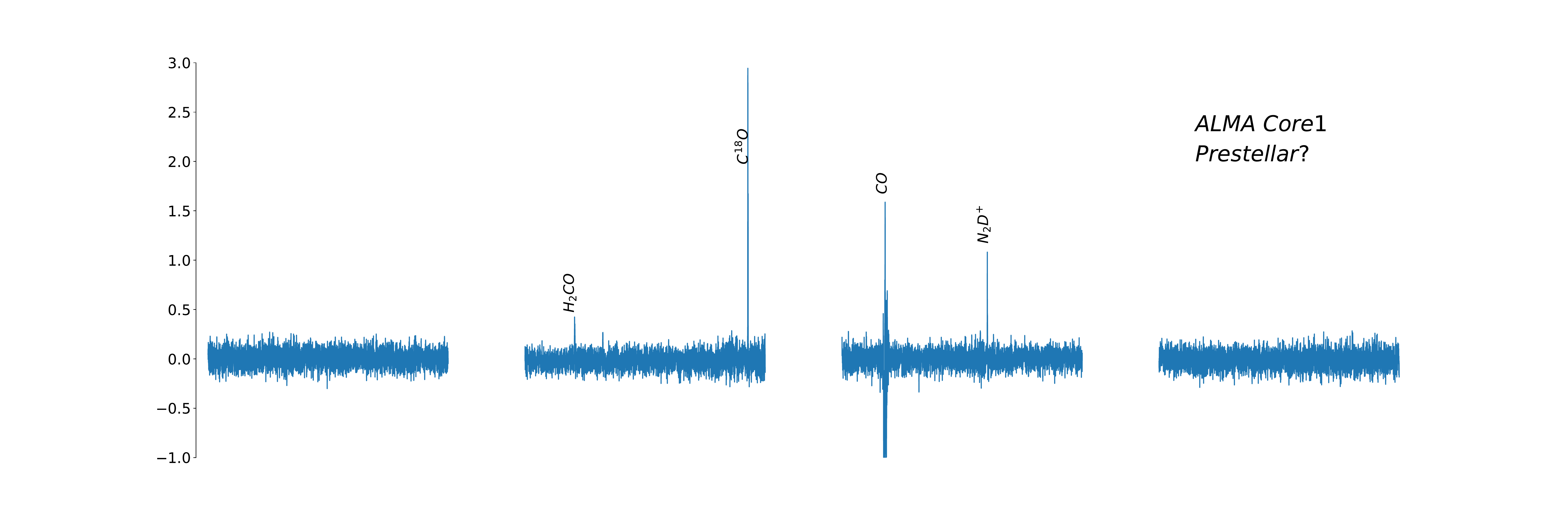}
            
            \includegraphics[width=1\textwidth]{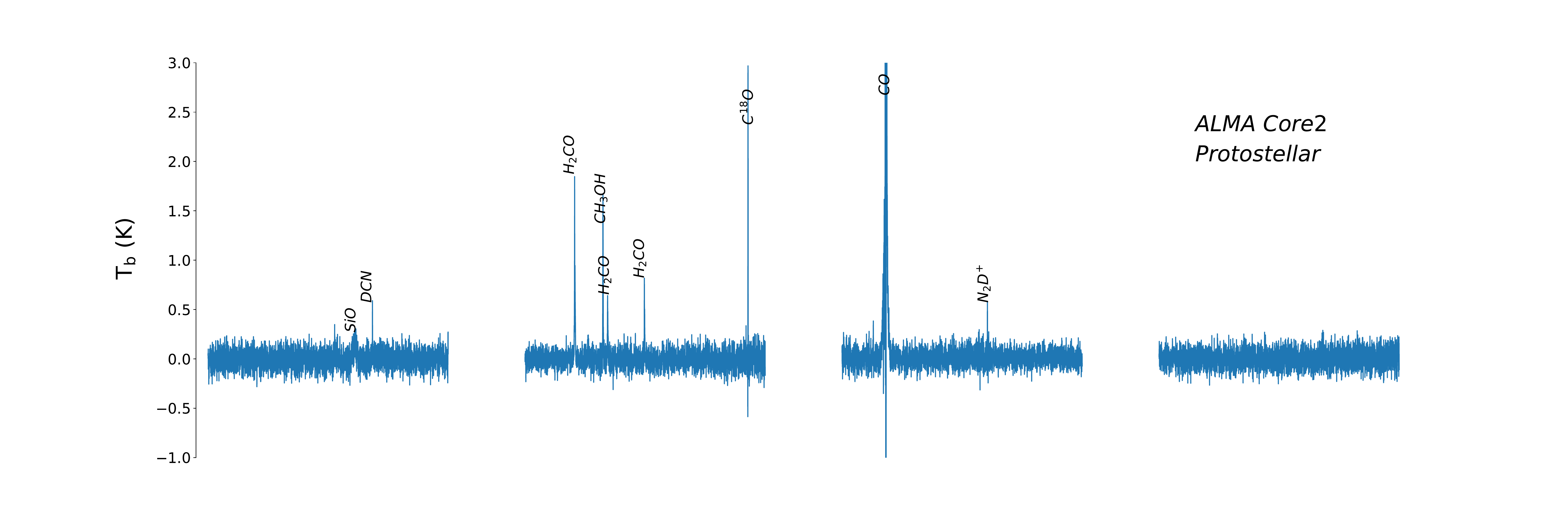}
            
            \includegraphics[width=1\textwidth]{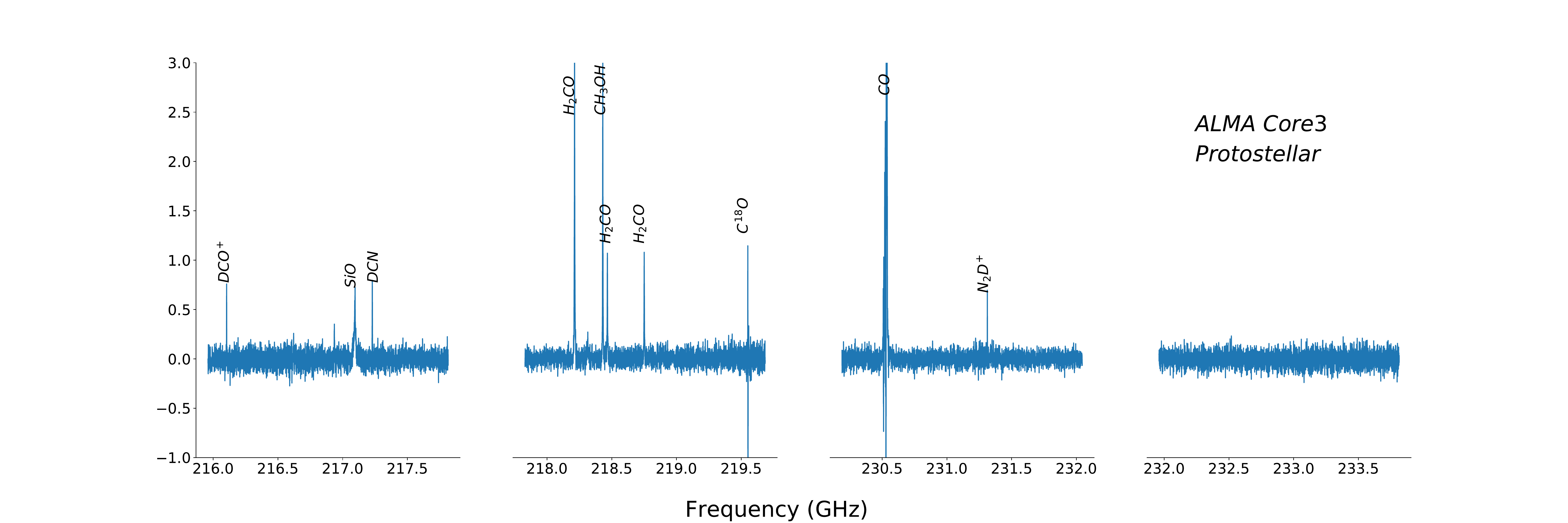}
    \caption{Spectra extracted from identified cores. The gaps separate the four ALMA spectral windows. Molecules with clear detection are labeled in the figure.}
    \label{Core Spectra}
\end{figure}

\begin{figure}[bt!]
    \centering
    \includegraphics[width=1.0\linewidth]{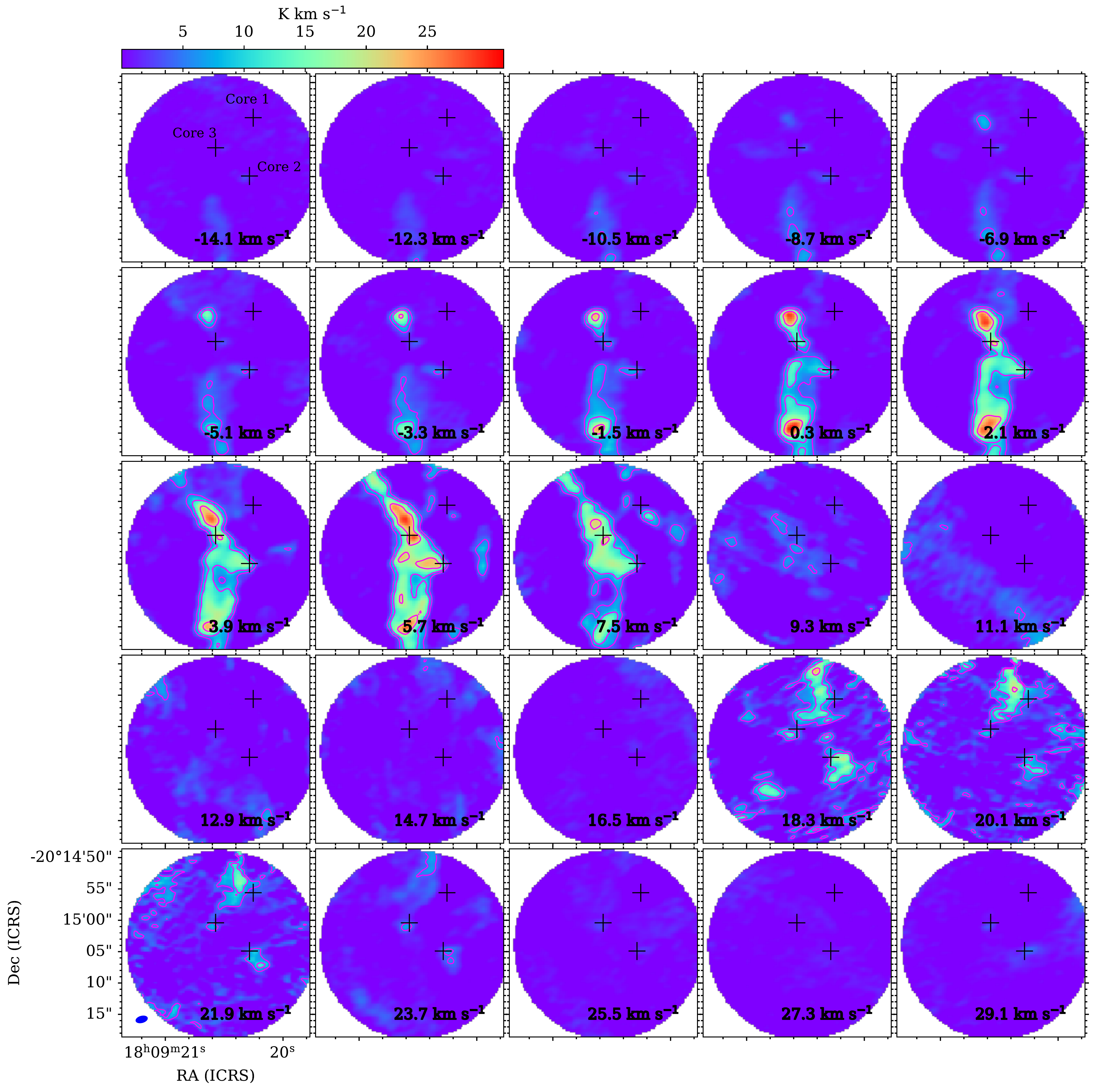}
    \caption{Channel maps of CO (2-1) after primary beam correction. The magenta contours show the integrated CO emission at levels of [6, 12, 24, 48, 96]$\sigma$, where $\sigma$ equals to 0.83 $\mathrm{K \ km \ s^{-1}}$. The black cross symbols represent the peak positions of compact cores. The left-bottom blue ellipse represents the beam size.}
    \label{fig:channel_map}
\end{figure}

\begin{figure}[bt!]
    \centering
    \includegraphics[width=1.0\linewidth]{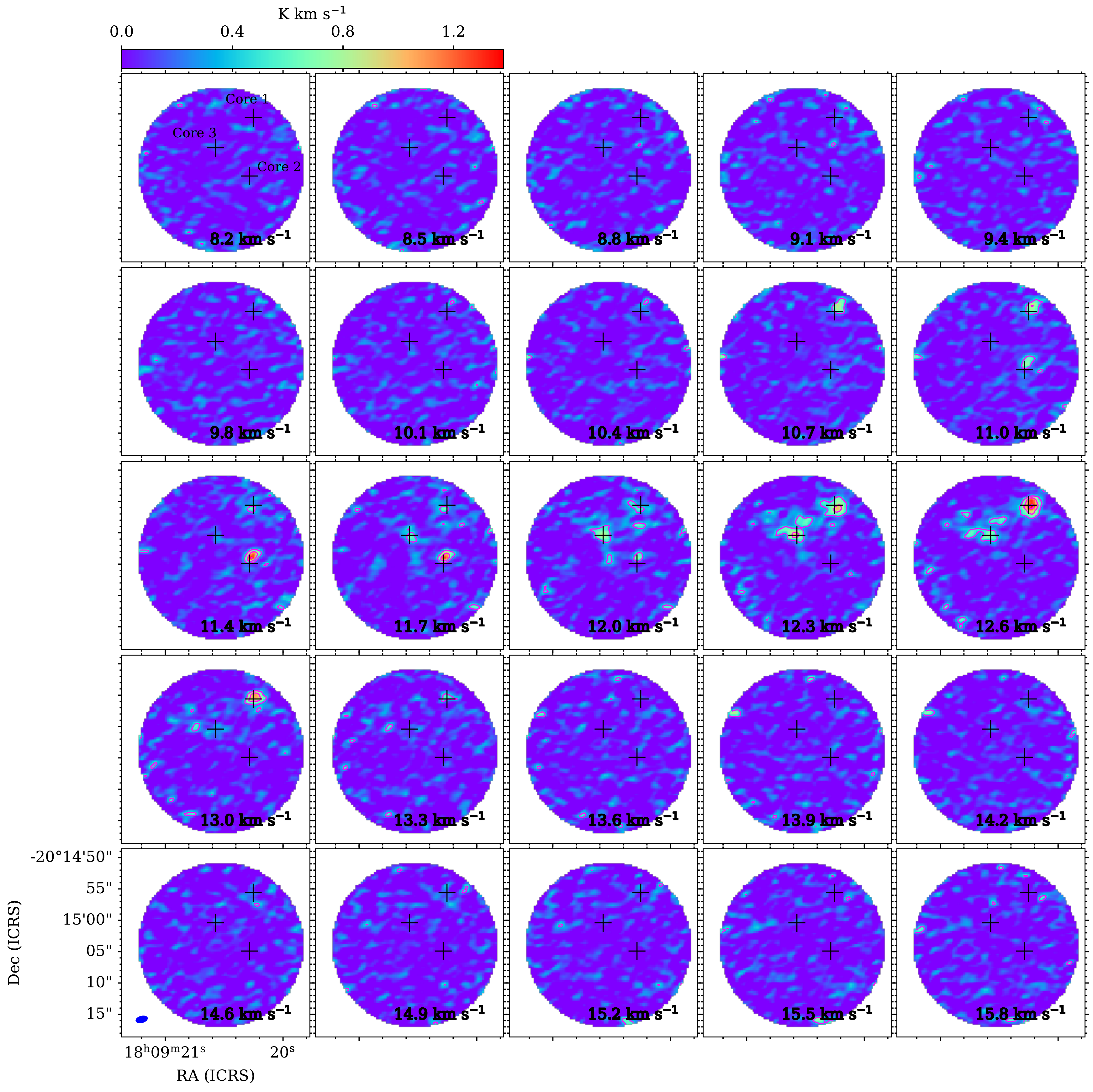}
    \caption{Channel maps of $\mathrm{N_2D^+}$ (3-2) after primary beam correction. The magenta contours show the $\mathrm{N_2D^+}$ emission at levels of [3, 6, 9, 12, 15]$\sigma$, where $\sigma$ equals to 0.15 $\mathrm{K \ km \ s^{-1}}$. The black cross symbols represent the peak positions of compact cores. The left-bottom blue ellipse represents the beam size.}
    \label{fig:n2dp_channel_map_pic}
\end{figure}

\appendix

\section{Spectra Extracted From Cores}
\label{sec:core_spectra}
\par Figure \ref{Core Spectra} shows the spectra extracted from identified cores. The typical warm gas tracers such as $\mathrm{CH_3OH}$ and $\mathrm{H_2CO}$ are detected in Core 2 and Core 3, indicating their protostellar properties. This deduction is consistent with the observations of outflows. No organic molecule emission is detected in Core 1, suggesting it at a very early evolutionary stage. Detailed discussion about the three deuterated molecules is described in Section \ref{deuterated molecules}. We find a clear trend in line richness from Core 1 to Core 3, changing from poor to abundant. Chemical difference is evident in the spectra of three cores, related to their different chemical ages.

\section{CO Outflow Identification}
\label{CO_outflow_identification}
\par The CO maps show complicated emission structures, which is difficult to identify outflow activities. We identify the CO outflows through carefully visual inspection. In order to detect the possible weak outflow lobes, we mask the strong environmental emission and check the channel maps around each core. Each identified lobe need to be detected over twenty channels ($\geq 7 \ \mathrm{km \ s^{-1}}$) with enough emission ($\geq 10 \ \mathrm{K \ km \ s^{-1}}$). For those lobes whose driving source is difficult to estimate, we judge them by the corresponding bipolar features. Besides the lobes that satisfy the criteria, we also find a weak blue lobe around Core 1 (o1a). Figure \ref{fig:channel_map} shows the CO channel maps from -15 $\mathrm{km \ s^{-1}}$ to 30 $\mathrm{km \ s^{-1}}$. Though the velocity range of this lobe is smaller than 10 $\mathrm{km \ s^{-1}}$, we still consider it a possible outflow lobe. Meanwhile, we can't exclude the possibility of the effect of side lobe or the extension of the blue lobe from o3a. In summary, this possible outflow association makes the nature of Core 1 unclear since it could be a prestellar core or a protostellar core driving an outflow at an earlier phase than Core 2 and 3.

\section{$\mathrm{N_2D^+}$ Channel Map}
\label{n2dp channel map}
\par As we observe multiple velocity components of $\mathrm{N_2D^+}$ in Core 1, we make a $\mathrm{N_2D^+}$ channel map to see whether the different velocity components are spatially resolved. Figure \ref{fig:n2dp_channel_map_pic} shows the $\mathrm{N_2D^+}$ channel maps from 8 $\mathrm{km \ s^{-1}}$ to 16 $\mathrm{km \ s^{-1}}$. All the velocity components are associated with the core. The two velocity components in Core 1 are not spatially resolved.


\bibliography{references} 

\begin{thebibliography}{}
\expandafter\ifx\csname natexlab\endcsname\relax\def\natexlab#1{#1}\fi
\providecommand{\url}[1]{\href{#1}{#1}}
\providecommand{\dodoi}[1]{doi:~\href{http://doi.org/#1}{\nolinkurl{#1}}}
\providecommand{\doeprint}[1]{\href{http://ascl.net/#1}{\nolinkurl{http://ascl.net/#1}}}
\providecommand{\doarXiv}[1]{\href{https://arxiv.org/abs/#1}{\nolinkurl{https://arxiv.org/abs/#1}}}

\bibitem[{{Aguirre} {et~al.}(2011){Aguirre}, {Ginsburg}, {Dunham}, {Drosback},
  {Bally}, {Battersby}, {Bradley}, {Cyganowski}, {Dowell}, {Evans}, {Glenn},
  {Harvey}, {Rosolowsky}, {Stringfellow}, {Walawender}, \&
  {Williams}}]{Aguirre_2011}
{Aguirre}, J.~E., {Ginsburg}, A.~G., {Dunham}, M.~K., {et~al.} 2011, \apjs,
  192, 4, \dodoi{10.1088/0067-0049/192/1/4}

\bibitem[{{Anderson} {et~al.}(2011){Anderson}, {Bania}, {Balser}, \&
  {Rood}}]{Anderson_2011}
{Anderson}, L.~D., {Bania}, T.~M., {Balser}, D.~S., \& {Rood}, R.~T. 2011,
  \apjs, 194, 32, \dodoi{10.1088/0067-0049/194/2/32}

\bibitem[{{Bacmann} {et~al.}(2003){Bacmann}, {Lefloch}, {Ceccarelli},
  {Steinacker}, {Castets}, \& {Loinard}}]{Bacmann_2003}
{Bacmann}, A., {Lefloch}, B., {Ceccarelli}, C., {et~al.} 2003, \apjl, 585, L55,
  \dodoi{10.1086/374263}

\bibitem[{{Barnes} {et~al.}(2021){Barnes}, {Henshaw}, {Fontani}, {Pineda},
  {Cosentino}, {Tan}, {Caselli}, {Jim{\'e}nez-Serra}, {Law}, {Avison},
  {Bigiel}, {Feng}, {Kong}, {Longmore}, {Moser}, {Parker}, {S{\'a}nchez-Monge},
  \& {Wang}}]{Barnes_2021}
{Barnes}, A.~T., {Henshaw}, J.~D., {Fontani}, F., {et~al.} 2021, \mnras, 503,
  4601, \dodoi{10.1093/mnras/stab803}

\bibitem[{{Baug} {et~al.}(2021){Baug}, {Wang}, {Liu}, {Wu}, {Li}, {Zhang},
  {Tang}, {Goldsmith}, {Liu}, {Tej}, {Bronfman}, {Kim}, {Li}, {Lee},
  {Tatematsu}, {Hirota}, \& {Toth}}]{Tapas_2021}
{Baug}, T., {Wang}, K., {Liu}, T., {et~al.} 2021, \mnras, 507, 4316,
  \dodoi{10.1093/mnras/stab1902}

\bibitem[{{Beltr{\'a}n} {et~al.}(2002){Beltr{\'a}n}, {Estalella}, {Ho},
  {Calvet}, {Anglada}, \& {Sep{\'u}lveda}}]{Beltran_2002}
{Beltr{\'a}n}, M.~T., {Estalella}, R., {Ho}, P. T.~P., {et~al.} 2002, \apj,
  565, 1069, \dodoi{10.1086/324683}

\bibitem[{{Bergin} \& {Tafalla}(2007)}]{Bergin_2007}
{Bergin}, E.~A., \& {Tafalla}, M. 2007, \araa, 45, 339,
  \dodoi{10.1146/annurev.astro.45.071206.100404}

\bibitem[{{Bertoldi} \& {McKee}(1992)}]{Bertoldi_1992}
{Bertoldi}, F., \& {McKee}, C.~F. 1992, \apj, 395, 140, \dodoi{10.1086/171638}

\bibitem[{{Beuther} {et~al.}(2019){Beuther}, {Ahmadi}, {Mottram}, {Linz},
  {Maud}, {Henning}, {Kuiper}, {Walsh}, {Johnston}, \&
  {Longmore}}]{Beuther_2019}
{Beuther}, H., {Ahmadi}, A., {Mottram}, J.~C., {et~al.} 2019, \aap, 621, A122,
  \dodoi{10.1051/0004-6361/201834064}

\bibitem[{{Beuther} {et~al.}(2021){Beuther}, {Gieser}, {Suri}, {Linz},
  {Klaassen}, {Semenov}, {Winters}, {Henning}, {Soler}, {Urquhart}, {Syed},
  {Feng}, {M{\"o}ller}, {Beltr{\'a}n}, {S{\'a}nchez-Monge}, {Longmore},
  {Peters}, {Ballesteros-Paredes}, {Schilke}, {Moscadelli}, {Palau},
  {Cesaroni}, {Lumsden}, {Pudritz}, {Wyrowski}, {Kuiper}, \&
  {Ahmadi}}]{Beuther_2021}
{Beuther}, H., {Gieser}, C., {Suri}, S., {et~al.} 2021, \aap, 649, A113,
  \dodoi{10.1051/0004-6361/202040106}

\bibitem[{{Bonnell} {et~al.}(2001){Bonnell}, {Bate}, {Clarke}, \&
  {Pringle}}]{Bonnell_2001}
{Bonnell}, I.~A., {Bate}, M.~R., {Clarke}, C.~J., \& {Pringle}, J.~E. 2001,
  \mnras, 323, 785, \dodoi{10.1046/j.1365-8711.2001.04270.x}

\bibitem[{{Bontemps} {et~al.}(1996){Bontemps}, {Andre}, {Terebey}, \&
  {Cabrit}}]{Bontemps_1996}
{Bontemps}, S., {Andre}, P., {Terebey}, S., \& {Cabrit}, S. 1996, \aap, 311,
  858

\bibitem[{{Butler} \& {Tan}(2012)}]{Bulter_2012}
{Butler}, M.~J., \& {Tan}, J.~C. 2012, \apj, 754, 5,
  \dodoi{10.1088/0004-637X/754/1/5}

\bibitem[{{Cabrit}(2009)}]{Cabrit_2009}
{Cabrit}, S. 2009, Astrophysics and Space Science Proceedings, 13, 247,
  \dodoi{10.1007/978-3-642-00576-3\_30}

\bibitem[{{Caselli} {et~al.}(2002){Caselli}, {Walmsley}, {Zucconi}, {Tafalla},
  {Dore}, \& {Myers}}]{Caselli_2002}
{Caselli}, P., {Walmsley}, C.~M., {Zucconi}, A., {et~al.} 2002, \apj, 565, 344,
  \dodoi{10.1086/324302}

\bibitem[{{Crapsi} {et~al.}(2005){Crapsi}, {Caselli}, {Walmsley}, {Myers},
  {Tafalla}, {Lee}, \& {Bourke}}]{Crapsi_2005}
{Crapsi}, A., {Caselli}, P., {Walmsley}, C.~M., {et~al.} 2005, \apj, 619, 379,
  \dodoi{10.1086/426472}

\bibitem[{{Csengeri} {et~al.}(2016){Csengeri}, {Leurini}, {Wyrowski},
  {Urquhart}, {Menten}, {Walmsley}, {Bontemps}, {Wienen}, {Beuther}, {Motte},
  {Nguyen-Luong}, {Schilke}, {Schuller}, {Zavagno}, \& {Sanna}}]{Csengeri_2016}
{Csengeri}, T., {Leurini}, S., {Wyrowski}, F., {et~al.} 2016, \aap, 586, A149,
  \dodoi{10.1051/0004-6361/201425404}

\bibitem[{{Dullemond} {et~al.}(2012){Dullemond}, {Juhasz}, {Pohl}, {Sereshti},
  {Shetty}, {Peters}, {Commercon}, \& {Flock}}]{Radmc-3d}
{Dullemond}, C.~P., {Juhasz}, A., {Pohl}, A., {et~al.} 2012, {RADMC-3D: A
  multi-purpose radiative transfer tool}.
\newblock \doeprint{1202.015}

\bibitem[{{Dunham} {et~al.}(2011){Dunham}, {Rosolowsky}, {Evans}, {Cyganowski},
  \& {Urquhart}}]{Dunham_2011}
{Dunham}, M.~K., {Rosolowsky}, E., {Evans}, Neal~J., I., {Cyganowski}, C., \&
  {Urquhart}, J.~S. 2011, \apj, 741, 110, \dodoi{10.1088/0004-637X/741/2/110}

\bibitem[{{Eden} {et~al.}(2017){Eden}, {Moore}, {Plume}, {Urquhart},
  {Thompson}, {Parsons}, {Dempsey}, {Rigby}, {Morgan}, {Thomas}, {Berry},
  {Buckle}, {Brunt}, {Butner}, {Carretero}, {Chrysostomou}, {Currie},
  {deVilliers}, {Fich}, {Gibb}, {Hoare}, {Jenness}, {Manser}, {Mottram},
  {Natario}, {Olguin}, {Peretto}, {Pestalozzi}, {Polychroni}, {Redman},
  {Salji}, {Summers}, {Tahani}, {Traficante}, {diFrancesco}, {Evans}, {Fuller},
  {Johnstone}, {Joncas}, {Longmore}, {Martin}, {Richer}, {Weferling}, {White},
  \& {Zhu}}]{Eden_2017}
{Eden}, D.~J., {Moore}, T.~J.~T., {Plume}, R., {et~al.} 2017, \mnras, 469,
  2163, \dodoi{10.1093/mnras/stx874}

\bibitem[{{Fontani} {et~al.}(2011){Fontani}, {Palau}, {Caselli},
  {S{\'a}nchez-Monge}, {Butler}, {Tan}, {Jim{\'e}nez-Serra}, {Busquet},
  {Leurini}, \& {Audard}}]{Fontani_2011}
{Fontani}, F., {Palau}, A., {Caselli}, P., {et~al.} 2011, \aap, 529, L7,
  \dodoi{10.1051/0004-6361/201116631}

\bibitem[{{Gerner} {et~al.}(2015){Gerner}, {Shirley}, {Beuther}, {Semenov},
  {Linz}, {Albertsson}, \& {Henning}}]{Gerner_2015}
{Gerner}, T., {Shirley}, Y.~L., {Beuther}, H., {et~al.} 2015, \aap, 579, A80,
  \dodoi{10.1051/0004-6361/201423989}

\bibitem[{{Gieser} {et~al.}(2021){Gieser}, {Beuther}, {Semenov}, {Ahmadi},
  {Suri}, {M{\"o}ller}, {Beltr{\'a}n}, {Klaassen}, {Zhang}, {Urquhart},
  {Henning}, {Feng}, {Galv{\'a}n-Madrid}, {de Souza Magalh{\~a}es},
  {Moscadelli}, {Longmore}, {Leurini}, {Kuiper}, {Peters}, {Menten},
  {Csengeri}, {Fuller}, {Wyrowski}, {Lumsden}, {S{\'a}nchez-Monge}, {Maud},
  {Linz}, {Palau}, {Schilke}, {Pety}, {Pudritz}, {Winters}, \&
  {Pi{\'e}tu}}]{Gieser_2021}
{Gieser}, C., {Beuther}, H., {Semenov}, D., {et~al.} 2021, \aap, 648, A66,
  \dodoi{10.1051/0004-6361/202039670}

\bibitem[{{Ginsburg} \& {Mirocha}(2011)}]{2011ascl.soft09001G}
{Ginsburg}, A., \& {Mirocha}, J. 2011, {PySpecKit: Python Spectroscopic
  Toolkit}.
\newblock \doeprint{1109.001}

\bibitem[{{G{\'o}mez-Ruiz} {et~al.}(2013){G{\'o}mez-Ruiz}, {Hirano}, {Leurini},
  \& {Liu}}]{Gomez_2013}
{G{\'o}mez-Ruiz}, A.~I., {Hirano}, N., {Leurini}, S., \& {Liu}, S.~Y. 2013,
  \aap, 558, A94, \dodoi{10.1051/0004-6361/201118473}

\bibitem[{{Green} {et~al.}(2011){Green}, {Evans}, {K{\'o}sp{\'a}l}, {van
  Kempen}, {Herczeg}, {Quanz}, {Henning}, {Lee}, {Dunham}, {Meeus}, {Bouwman},
  {van Dishoeck}, {Chen}, {G{\"u}del}, {Skinner}, {Merello}, {Pooley},
  {Rebull}, \& {Guieu}}]{Green_2011}
{Green}, J.~D., {Evans}, Neal~J., I., {K{\'o}sp{\'a}l}, {\'A}., {et~al.} 2011,
  \apjl, 731, L25, \dodoi{10.1088/2041-8205/731/2/L25}

\bibitem[{{Hartmann} {et~al.}(2012){Hartmann}, {Ballesteros-Paredes}, \&
  {Heitsch}}]{Hartmann_2012}
{Hartmann}, L., {Ballesteros-Paredes}, J., \& {Heitsch}, F. 2012, \mnras, 420,
  1457, \dodoi{10.1111/j.1365-2966.2011.20131.x}

\bibitem[{{Hildebrand}(1983)}]{Hildebrand_1983}
{Hildebrand}, R.~H. 1983, \qjras, 24, 267

\bibitem[{{Jeans}(1902)}]{Jeans_1902}
{Jeans}, J.~H. 1902, Philosophical Transactions of the Royal Society of London
  Series A, 199, 1, \dodoi{10.1098/rsta.1902.0012}

\bibitem[{{Kauffmann} {et~al.}(2008){Kauffmann}, {Bertoldi}, {Bourke}, {Evans},
  \& {Lee}}]{Kauffmann_2008}
{Kauffmann}, J., {Bertoldi}, F., {Bourke}, T.~L., {Evans}, N.~J., I., \& {Lee},
  C.~W. 2008, \aap, 487, 993, \dodoi{10.1051/0004-6361:200809481}

\bibitem[{{Kauffmann} \& {Pillai}(2010)}]{Kauffmann_2010}
{Kauffmann}, J., \& {Pillai}, T. 2010, \apjl, 723, L7,
  \dodoi{10.1088/2041-8205/723/1/L7}

\bibitem[{{Kong} {et~al.}(2021){Kong}, {Arce}, {Shirley}, \&
  {Glasgow}}]{Kong_2021}
{Kong}, S., {Arce}, H.~G., {Shirley}, Y., \& {Glasgow}, C. 2021, \apj, 912,
  156, \dodoi{10.3847/1538-4357/abefe7}

\bibitem[{{Kong} {et~al.}(2015){Kong}, {Caselli}, {Tan}, {Wakelam}, \&
  {Sipil{\"a}}}]{Kong_2015}
{Kong}, S., {Caselli}, P., {Tan}, J.~C., {Wakelam}, V., \& {Sipil{\"a}}, O.
  2015, \apj, 804, 98, \dodoi{10.1088/0004-637X/804/2/98}

\bibitem[{{Kong} {et~al.}(2017){Kong}, {Tan}, {Caselli}, {Fontani}, {Liu}, \&
  {Butler}}]{Kong_2017}
{Kong}, S., {Tan}, J.~C., {Caselli}, P., {et~al.} 2017, \apj, 834, 193,
  \dodoi{10.3847/1538-4357/834/2/193}

\bibitem[{{Konigl} \& {Pudritz}(2000)}]{Konigl_2000}
{Konigl}, A., \& {Pudritz}, R.~E. 2000, in Protostars and Planets IV, ed.
  V.~{Mannings}, A.~P. {Boss}, \& S.~S. {Russell}, 759

\bibitem[{{Kroupa}(2001)}]{Kroupa_2001}
{Kroupa}, P. 2001, \mnras, 322, 231, \dodoi{10.1046/j.1365-8711.2001.04022.x}

\bibitem[{{Larson}(2003)}]{Larson_2003}
{Larson}, R.~B. 2003, in Astronomical Society of the Pacific Conference Series,
  Vol. 287, Galactic Star Formation Across the Stellar Mass Spectrum, ed. J.~M.
  {De Buizer} \& N.~S. {van der Bliek}, 65--80

\bibitem[{{Li} {et~al.}(2013){Li}, {Kauffmann}, {Zhang}, \& {Chen}}]{Li_2013}
{Li}, D., {Kauffmann}, J., {Zhang}, Q., \& {Chen}, W. 2013, \apjl, 768, L5,
  \dodoi{10.1088/2041-8205/768/1/L5}

\bibitem[{{Li} {et~al.}(2019{\natexlab{a}}){Li}, {Zhang}, {Pillai}, {Stephens},
  {Wang}, \& {Li}}]{Li_2019_b}
{Li}, S., {Zhang}, Q., {Pillai}, T., {et~al.} 2019{\natexlab{a}}, \apj, 886,
  130, \dodoi{10.3847/1538-4357/ab464e}

\bibitem[{{Li} {et~al.}(2019{\natexlab{b}}){Li}, {Wang}, {Fang}, {Zhang}, {Li},
  {Zhang}, {Li}, {Zhu}, \& {Zeng}}]{Li_2019_a}
{Li}, S., {Wang}, J., {Fang}, M., {et~al.} 2019{\natexlab{b}}, \apj, 878, 29,
  \dodoi{10.3847/1538-4357/ab1e4c}

\bibitem[{{Li} {et~al.}(2021){Li}, {Lu}, {Zhang}, {Lee}, {Sanhueza}, {Beuther},
  {Jim{\'e}nez-Serra}, {Qiu}, {Palau}, {Feng}, {Pillai}, {Kim}, {Liu},
  {Girart}, {Liu}, {Wang}, {Wang}, {Liu}, {Smith}, {Li}, {Lee}, {Li}, {Li},
  {Kim}, {Yue}, \& {Strom}}]{Li_2021}
{Li}, S., {Lu}, X., {Zhang}, Q., {et~al.} 2021, \apjl, 912, L7,
  \dodoi{10.3847/2041-8213/abf64f}

\bibitem[{{Liu} {et~al.}(2017){Liu}, {Lacy}, {Li}, {Wang}, {Qin}, {Zhang},
  {Kim}, {Garay}, {Wu}, {Mardones}, {Zhu}, {Tatematsu}, {Hirota}, {Ren}, {Liu},
  {Chen}, {Su}, \& {Li}}]{Liu_2017}
{Liu}, T., {Lacy}, J., {Li}, P.~S., {et~al.} 2017, \apj, 849, 25,
  \dodoi{10.3847/1538-4357/aa8d73}

\bibitem[{{Lu} {et~al.}(2017){Lu}, {Zhang}, {Kauffmann}, {Pillai}, {Longmore},
  {Kruijssen}, {Battersby}, {Liu}, {Ginsburg}, {Mills}, {Zhang}, \&
  {Gu}}]{Lu_2017}
{Lu}, X., {Zhang}, Q., {Kauffmann}, J., {et~al.} 2017, \apj, 839, 1,
  \dodoi{10.3847/1538-4357/aa67f7}

\bibitem[{{Lu} {et~al.}(2018){Lu}, {Zhang}, {Liu}, {Sanhueza}, {Tatematsu},
  {Feng}, {Smith}, {Myers}, {Sridharan}, \& {Gu}}]{Lu_2018}
{Lu}, X., {Zhang}, Q., {Liu}, H.~B., {et~al.} 2018, \apj, 855, 9,
  \dodoi{10.3847/1538-4357/aaad11}

\bibitem[{{Lu} {et~al.}(2021){Lu}, {Li}, {Ginsburg}, {Longmore}, {Kruijssen},
  {Walker}, {Feng}, {Zhang}, {Battersby}, {Pillai}, {Mills}, {Kauffmann},
  {Cheng}, \& {Inutsuka}}]{Lu_2021}
{Lu}, X., {Li}, S., {Ginsburg}, A., {et~al.} 2021, \apj, 909, 177,
  \dodoi{10.3847/1538-4357/abde3c}

\bibitem[{{MacLaren} {et~al.}(1988){MacLaren}, {Richardson}, \&
  {Wolfendale}}]{Maclaren_1988}
{MacLaren}, I., {Richardson}, K.~M., \& {Wolfendale}, A.~W. 1988, \apj, 333,
  821, \dodoi{10.1086/166791}

\bibitem[{{Mangum} \& {Shirley}(2015)}]{Mangum&Shirley_2015}
{Mangum}, J.~G., \& {Shirley}, Y.~L. 2015, \pasp, 127, 266,
  \dodoi{10.1086/680323}

\bibitem[{{Mangum} \& {Wootten}(1993)}]{Mangum_1993}
{Mangum}, J.~G., \& {Wootten}, A. 1993, \apjs, 89, 123, \dodoi{10.1086/191841}

\bibitem[{{Maud} {et~al.}(2018){Maud}, {Cesaroni}, {Kumar}, {van der Tak},
  {Allen}, {Hoare}, {Klaassen}, {Harsono}, {Hogerheijde}, {S{\'a}nchez-Monge},
  {Schilke}, {Ahmadi}, {Beltr{\'a}n}, {Beuther}, {Csengeri}, {Etoka}, {Fuller},
  {Galv{\'a}n-Madrid}, {Goddi}, {Henning}, {Johnston}, {Kuiper}, {Lumsden},
  {Moscadelli}, {Mottram}, {Peters}, {Rivilla}, {Testi}, {Vig}, {de Wit}, \&
  {Zinnecker}}]{Maud_2018}
{Maud}, L.~T., {Cesaroni}, R., {Kumar}, M.~S.~N., {et~al.} 2018, \aap, 620,
  A31, \dodoi{10.1051/0004-6361/201833908}

\bibitem[{{McKee} \& {Tan}(2003)}]{McKee_2003}
{McKee}, C.~F., \& {Tan}, J.~C. 2003, \apj, 585, 850, \dodoi{10.1086/346149}

\bibitem[{{McMullin} {et~al.}(2007){McMullin}, {Waters}, {Schiebel}, {Young},
  \& {Golap}}]{McMullin_2007}
{McMullin}, J.~P., {Waters}, B., {Schiebel}, D., {Young}, W., \& {Golap}, K.
  2007, in Astronomical Society of the Pacific Conference Series, Vol. 376,
  Astronomical Data Analysis Software and Systems XVI, ed. R.~A. {Shaw},
  F.~{Hill}, \& D.~J. {Bell}, 127

\bibitem[{{Molinari} {et~al.}(2010){Molinari}, {Swinyard}, {Bally}, {Barlow},
  {Bernard}, {Martin}, {Moore}, {Noriega-Crespo}, {Plume}, {Testi}, {Zavagno},
  {Abergel}, {Ali}, {Andr{\'e}}, {Baluteau}, {Benedettini}, {Bern{\'e}},
  {Billot}, {Blommaert}, {Bontemps}, {Boulanger}, {Brand}, {Brunt}, {Burton},
  {Campeggio}, {Carey}, {Caselli}, {Cesaroni}, {Cernicharo}, {Chakrabarti},
  {Chrysostomou}, {Codella}, {Cohen}, {Compiegne}, {Davis}, {de Bernardis}, {de
  Gasperis}, {Di Francesco}, {di Giorgio}, {Elia}, {Faustini}, {Fischera},
  {Fukui}, {Fuller}, {Ganga}, {Garcia-Lario}, {Giard}, {Giardino}, {Glenn},
  {Goldsmith}, {Griffin}, {Hoare}, {Huang}, {Jiang}, {Joblin}, {Joncas},
  {Juvela}, {Kirk}, {Lagache}, {Li}, {Lim}, {Lord}, {Lucas}, {Maiolo},
  {Marengo}, {Marshall}, {Masi}, {Massi}, {Matsuura}, {Meny}, {Minier},
  {Miville-Desch{\^e}nes}, {Montier}, {Motte}, {M{\"u}ller}, {Natoli}, {Neves},
  {Olmi}, {Paladini}, {Paradis}, {Pestalozzi}, {Pezzuto}, {Piacentini},
  {Pomar{\`e}s}, {Popescu}, {Reach}, {Richer}, {Ristorcelli}, {Roy}, {Royer},
  {Russeil}, {Saraceno}, {Sauvage}, {Schilke}, {Schneider-Bontemps},
  {Schuller}, {Schultz}, {Shepherd}, {Sibthorpe}, {Smith}, {Smith},
  {Spinoglio}, {Stamatellos}, {Strafella}, {Stringfellow}, {Sturm}, {Taylor},
  {Thompson}, {Tuffs}, {Umana}, {Valenziano}, {Vavrek}, {Viti}, {Waelkens},
  {Ward-Thompson}, {White}, {Wyrowski}, {Yorke}, \& {Zhang}}]{Molinari_2010}
{Molinari}, S., {Swinyard}, B., {Bally}, J., {et~al.} 2010, \pasp, 122, 314,
  \dodoi{10.1086/651314}

\bibitem[{{M{\"o}ller} {et~al.}(2017){M{\"o}ller}, {Endres}, \&
  {Schilke}}]{Moller_2017}
{M{\"o}ller}, T., {Endres}, C., \& {Schilke}, P. 2017, \aap, 598, A7,
  \dodoi{10.1051/0004-6361/201527203}

\bibitem[{{Morii} {et~al.}(2021){Morii}, {Sanhueza}, {Nakamura}, {Jackson},
  {Li}, {Beuther}, {Zhang}, {Feng}, {Tafoya}, {Guzm{\'a}n}, {Izumi}, {Sakai},
  {Lu}, {Tatematsu}, {Ohashi}, {Silva}, {Olguin}, \& {Contreras}}]{Morii_2021}
{Morii}, K., {Sanhueza}, P., {Nakamura}, F., {et~al.} 2021, \apj, 923, 147,
  \dodoi{10.3847/1538-4357/ac2365}

\bibitem[{{Motte} {et~al.}(2018){Motte}, {Bontemps}, \& {Louvet}}]{Motte_2018}
{Motte}, F., {Bontemps}, S., \& {Louvet}, F. 2018, \araa, 56, 41,
  \dodoi{10.1146/annurev-astro-091916-055235}

\bibitem[{{Ossenkopf} \& {Henning}(1994)}]{OH94}
{Ossenkopf}, V., \& {Henning}, T. 1994, \aap, 291, 943

\bibitem[{{Padoan} {et~al.}(2020){Padoan}, {Pan}, {Juvela}, {Haugb{\o}lle}, \&
  {Nordlund}}]{Padoan_2020}
{Padoan}, P., {Pan}, L., {Juvela}, M., {Haugb{\o}lle}, T., \& {Nordlund},
  {\r{A}}. 2020, \apj, 900, 82, \dodoi{10.3847/1538-4357/abaa47}

\bibitem[{{Palau} {et~al.}(2014){Palau}, {Estalella}, {Girart}, {Fuente},
  {Fontani}, {Commer{\c{c}}on}, {Busquet}, {Bontemps}, {S{\'a}nchez-Monge},
  {Zapata}, {Zhang}, {Hennebelle}, \& {di Francesco}}]{Palau_2014}
{Palau}, A., {Estalella}, R., {Girart}, J.~M., {et~al.} 2014, \apj, 785, 42,
  \dodoi{10.1088/0004-637X/785/1/42}

\bibitem[{{Palau} {et~al.}(2018){Palau}, {Zapata}, {Rom{\'a}n-Z{\'u}{\~n}iga},
  {S{\'a}nchez-Monge}, {Estalella}, {Busquet}, {Girart}, {Fuente}, \&
  {Commer{\c{c}}on}}]{Palau_2018}
{Palau}, A., {Zapata}, L.~A., {Rom{\'a}n-Z{\'u}{\~n}iga}, C.~G., {et~al.} 2018,
  \apj, 855, 24, \dodoi{10.3847/1538-4357/aaad03}

\bibitem[{{Phan-Bao} {et~al.}(2014){Phan-Bao}, {Lee}, {Ho}, {Dang-Duc}, \&
  {Li}}]{Bao_2014}
{Phan-Bao}, N., {Lee}, C.-F., {Ho}, P. T.~P., {Dang-Duc}, C., \& {Li}, D. 2014,
  \apj, 795, 70, \dodoi{10.1088/0004-637X/795/1/70}

\bibitem[{{Pillai} {et~al.}(2011){Pillai}, {Kauffmann}, {Wyrowski}, {Hatchell},
  {Gibb}, \& {Thompson}}]{Pillai_2011}
{Pillai}, T., {Kauffmann}, J., {Wyrowski}, F., {et~al.} 2011, \aap, 530, A118,
  \dodoi{10.1051/0004-6361/201015899}

\bibitem[{{Pillai} {et~al.}(2019){Pillai}, {Kauffmann}, {Zhang}, {Sanhueza},
  {Leurini}, {Wang}, {Sridharan}, \& {K{\"o}nig}}]{Pillai_2019}
{Pillai}, T., {Kauffmann}, J., {Zhang}, Q., {et~al.} 2019, \aap, 622, A54,
  \dodoi{10.1051/0004-6361/201732570}

\bibitem[{{Pillai} {et~al.}(2007){Pillai}, {Wyrowski}, {Hatchell}, {Gibb}, \&
  {Thompson}}]{Pillai_2007}
{Pillai}, T., {Wyrowski}, F., {Hatchell}, J., {Gibb}, A.~G., \& {Thompson},
  M.~A. 2007, \aap, 467, 207, \dodoi{10.1051/0004-6361:20065682}

\bibitem[{{Qiu} {et~al.}(2009){Qiu}, {Zhang}, {Wu}, \& {Chen}}]{Qiu_2009}
{Qiu}, K., {Zhang}, Q., {Wu}, J., \& {Chen}, H.-R. 2009, \apj, 696, 66,
  \dodoi{10.1088/0004-637X/696/1/66}

\bibitem[{{Sakai} {et~al.}(2021){Sakai}, {Sanhueza}, {Furuya}, {Tatematsu},
  {Li}, {Aikawa}, {Lu}, {Zhang}, {Morii}, {Nakamura}, {Takemura}, {Izumi},
  {Hirota}, {Silva}, {Guzm{\'a}n}, {Sakai}, \& {Yamamoto}}]{Sakai_2021}
{Sakai}, T., {Sanhueza}, P., {Furuya}, K., {et~al.} 2021, arXiv e-prints,
  arXiv:2111.13325.
\newblock \doarXiv{2111.13325}

\bibitem[{{S{\'a}nchez-Monge} {et~al.}(2013){S{\'a}nchez-Monge}, {Palau},
  {Fontani}, {Busquet}, {Ju{\'a}rez}, {Estalella}, {Tan}, {Sep{\'u}lveda},
  {Ho}, {Zhang}, \& {Kurtz}}]{Sanchez-Monge_2013}
{S{\'a}nchez-Monge}, {\'A}., {Palau}, A., {Fontani}, F., {et~al.} 2013, \mnras,
  432, 3288, \dodoi{10.1093/mnras/stt679}

\bibitem[{{Sanhueza} {et~al.}(2013){Sanhueza}, {Jackson}, {Foster},
  {Jimenez-Serra}, {Dirienzo}, \& {Pillai}}]{Sanhueza_2013}
{Sanhueza}, P., {Jackson}, J.~M., {Foster}, J.~B., {et~al.} 2013, \apj, 773,
  123, \dodoi{10.1088/0004-637X/773/2/123}

\bibitem[{{Sanhueza} {et~al.}(2017){Sanhueza}, {Jackson}, {Zhang},
  {Guzm{\'a}n}, {Lu}, {Stephens}, {Wang}, \& {Tatematsu}}]{Sanhueza_2017}
{Sanhueza}, P., {Jackson}, J.~M., {Zhang}, Q., {et~al.} 2017, \apj, 841, 97,
  \dodoi{10.3847/1538-4357/aa6ff8}

\bibitem[{{Sanhueza} {et~al.}(2019){Sanhueza}, {Contreras}, {Wu}, {Jackson},
  {Guzm{\'a}n}, {Zhang}, {Li}, {Lu}, {Silva}, {Izumi}, {Liu}, {Miura},
  {Tatematsu}, {Sakai}, {Beuther}, {Garay}, {Ohashi}, {Saito}, {Nakamura},
  {Saigo}, {Veena}, {Nguyen-Luong}, \& {Tafoya}}]{Sanhueza_2019}
{Sanhueza}, P., {Contreras}, Y., {Wu}, B., {et~al.} 2019, \apj, 886, 102,
  \dodoi{10.3847/1538-4357/ab45e9}

\bibitem[{{Santamar{\'\i}a-Miranda} {et~al.}(2020){Santamar{\'\i}a-Miranda},
  {de Gregorio-Monsalvo}, {Hu{\'e}lamo}, {Plunkett}, {Ribas}, {Comer{\'o}n},
  {Schreiber}, {L{\'o}pez}, {Mu{\v{z}}i{\'c}}, \& {Testi}}]{Miranda_2020}
{Santamar{\'\i}a-Miranda}, A., {de Gregorio-Monsalvo}, I., {Hu{\'e}lamo}, N.,
  {et~al.} 2020, \aap, 640, A13, \dodoi{10.1051/0004-6361/202038128}

\bibitem[{{Schuller} {et~al.}(2009){Schuller}, {Menten}, {Contreras},
  {Wyrowski}, {Schilke}, {Bronfman}, {Henning}, {Walmsley}, {Beuther},
  {Bontemps}, {Cesaroni}, {Deharveng}, {Garay}, {Herpin}, {Lefloch}, {Linz},
  {Mardones}, {Minier}, {Molinari}, {Motte}, {Nyman}, {Reveret}, {Risacher},
  {Russeil}, {Schneider}, {Testi}, {Troost}, {Vasyunina}, {Wienen}, {Zavagno},
  {Kovacs}, {Kreysa}, {Siringo}, \& {Wei{\ss}}}]{Schuller_2009}
{Schuller}, F., {Menten}, K.~M., {Contreras}, Y., {et~al.} 2009, \aap, 504,
  415, \dodoi{10.1051/0004-6361/200811568}

\bibitem[{{Scoville} \& {Kwan}(1976)}]{Scoville_1976}
{Scoville}, N.~Z., \& {Kwan}, J. 1976, \apj, 206, 718, \dodoi{10.1086/154432}

\bibitem[{{Smith} {et~al.}(2009){Smith}, {Longmore}, \& {Bonnell}}]{Smith_2009}
{Smith}, R.~J., {Longmore}, S., \& {Bonnell}, I. 2009, \mnras, 400, 1775,
  \dodoi{10.1111/j.1365-2966.2009.15621.x}

\bibitem[{{Svoboda} {et~al.}(2016){Svoboda}, {Shirley}, {Battersby},
  {Rosolowsky}, {Ginsburg}, {Ellsworth-Bowers}, {Pestalozzi}, {Dunham},
  {Evans}, {Bally}, \& {Glenn}}]{Svoboda_2016}
{Svoboda}, B.~E., {Shirley}, Y.~L., {Battersby}, C., {et~al.} 2016, \apj, 822,
  59, \dodoi{10.3847/0004-637X/822/2/59}

\bibitem[{{Svoboda} {et~al.}(2019){Svoboda}, {Shirley}, {Traficante},
  {Battersby}, {Fuller}, {Zhang}, {Beuther}, {Peretto}, {Brogan}, \&
  {Hunter}}]{Svoboda_2019}
{Svoboda}, B.~E., {Shirley}, Y.~L., {Traficante}, A., {et~al.} 2019, \apj, 886,
  36, \dodoi{10.3847/1538-4357/ab40ca}

\bibitem[{{Tackenberg} {et~al.}(2012){Tackenberg}, {Beuther}, {Henning},
  {Schuller}, {Wienen}, {Motte}, {Wyrowski}, {Bontemps}, {Bronfman}, {Menten},
  {Testi}, \& {Lefloch}}]{Tackenberg_2012}
{Tackenberg}, J., {Beuther}, H., {Henning}, T., {et~al.} 2012, \aap, 540, A113,
  \dodoi{10.1051/0004-6361/201117412}

\bibitem[{{Tan} {et~al.}(2014){Tan}, {Beltr{\'a}n}, {Caselli}, {Fontani},
  {Fuente}, {Krumholz}, {McKee}, \& {Stolte}}]{Tan_2014}
{Tan}, J.~C., {Beltr{\'a}n}, M.~T., {Caselli}, P., {et~al.} 2014, in Protostars
  and Planets VI, ed. H.~{Beuther}, R.~S. {Klessen}, C.~P. {Dullemond}, \&
  T.~{Henning}, 149

\bibitem[{{Tang} {et~al.}(2021){Tang}, {Palau}, {Zapata}, \& {Qin}}]{Tang_2021}
{Tang}, M., {Palau}, A., {Zapata}, L.~A., \& {Qin}, S.-L. 2021, arXiv e-prints,
  arXiv:2109.07658.
\newblock \doarXiv{2109.07658}

\bibitem[{{Traficante} {et~al.}(2015){Traficante}, {Fuller}, {Peretto},
  {Pineda}, \& {Molinari}}]{Traficante_2015}
{Traficante}, A., {Fuller}, G.~A., {Peretto}, N., {Pineda}, J.~E., \&
  {Molinari}, S. 2015, \mnras, 451, 3089, \dodoi{10.1093/mnras/stv1158}

\bibitem[{{Tychoniec} {et~al.}(2019){Tychoniec}, {Hull}, {Kristensen}, {Tobin},
  {Le Gouellec}, \& {van Dishoeck}}]{tychoniec_2019}
{Tychoniec}, {\L}., {Hull}, C. L.~H., {Kristensen}, L.~E., {et~al.} 2019, \aap,
  632, A101, \dodoi{10.1051/0004-6361/201935409}

\bibitem[{{Urquhart} {et~al.}(2013){Urquhart}, {Moore}, {Schuller}, {Wyrowski},
  {Menten}, {Thompson}, {Csengeri}, {Walmsley}, {Bronfman}, \&
  {K{\"o}nig}}]{Urquhart_2013}
{Urquhart}, J.~S., {Moore}, T.~J.~T., {Schuller}, F., {et~al.} 2013, \mnras,
  431, 1752, \dodoi{10.1093/mnras/stt287}

\bibitem[{{Wang} {et~al.}(2011){Wang}, {Zhang}, {Wu}, \& {Zhang}}]{Wang_2011}
{Wang}, K., {Zhang}, Q., {Wu}, Y., \& {Zhang}, H. 2011, \apj, 735, 64,
  \dodoi{10.1088/0004-637X/735/1/64}

\bibitem[{{Wang} {et~al.}(2014){Wang}, {Zhang}, {Testi}, {van der Tak}, {Wu},
  {Zhang}, {Pillai}, {Wyrowski}, {Carey}, {Ragan}, \& {Henning}}]{Wang_2014}
{Wang}, K., {Zhang}, Q., {Testi}, L., {et~al.} 2014, \mnras, 439, 3275,
  \dodoi{10.1093/mnras/stu127}

\bibitem[{{Wang} {et~al.}(2010){Wang}, {Li}, {Abel}, \& {Nakamura}}]{Wang_2010}
{Wang}, P., {Li}, Z.-Y., {Abel}, T., \& {Nakamura}, F. 2010, \apj, 709, 27,
  \dodoi{10.1088/0004-637X/709/1/27}

\bibitem[{{Wienen} {et~al.}(2012){Wienen}, {Wyrowski}, {Schuller}, {Menten},
  {Walmsley}, {Bronfman}, \& {Motte}}]{Wienen_2012}
{Wienen}, M., {Wyrowski}, F., {Schuller}, F., {et~al.} 2012, \aap, 544, A146,
  \dodoi{10.1051/0004-6361/201118107}

\bibitem[{{Yuan} {et~al.}(2017){Yuan}, {Wu}, {Ellingsen}, {Evans}, {Henkel},
  {Wang}, {Liu}, {Liu}, {Li}, \& {Zavagno}}]{Yuan_2017}
{Yuan}, J., {Wu}, Y., {Ellingsen}, S.~P., {et~al.} 2017, \apjs, 231, 11,
  \dodoi{10.3847/1538-4365/aa7204}

\bibitem[{{Zhang} {et~al.}(2005){Zhang}, {Hunter}, {Brand}, {Sridharan},
  {Cesaroni}, {Molinari}, {Wang}, \& {Kramer}}]{Zhang_2005}
{Zhang}, Q., {Hunter}, T.~R., {Brand}, J., {et~al.} 2005, \apj, 625, 864,
  \dodoi{10.1086/429660}

\bibitem[{{Zhang} \& {Wang}(2011)}]{Zhang_2011}
{Zhang}, Q., \& {Wang}, K. 2011, \apj, 733, 26,
  \dodoi{10.1088/0004-637X/733/1/26}

\bibitem[{{Zhang} {et~al.}(2015){Zhang}, {Wang}, {Lu}, \&
  {Jim{\'e}nez-Serra}}]{Zhang_2015}
{Zhang}, Q., {Wang}, K., {Lu}, X., \& {Jim{\'e}nez-Serra}, I. 2015, \apj, 804,
  141, \dodoi{10.1088/0004-637X/804/2/141}

\bibitem[{{Zhang} {et~al.}(2009){Zhang}, {Wang}, {Pillai}, \&
  {Rathborne}}]{Zhang_2009}
{Zhang}, Q., {Wang}, Y., {Pillai}, T., \& {Rathborne}, J. 2009, \apj, 696, 268,
  \dodoi{10.1088/0004-637X/696/1/268}

\bibitem[{{Zhang} {et~al.}(2021){Zhang}, {Zavagno}, {L{\'o}pez-Sepulcre},
  {Liu}, {Louvet}, {Figueira}, {Russeil}, {Wu}, {Yuan}, \&
  {Pillai}}]{Zhang_2021}
{Zhang}, S., {Zavagno}, A., {L{\'o}pez-Sepulcre}, A., {et~al.} 2021, \aap, 646,
  A25, \dodoi{10.1051/0004-6361/202038421}

\end{thebibliography}
\bibliographystyle{aasjournal}

\label{lastpage}
\end{document}